\documentclass[conference, nofonttune]{IEEEtran}

\usepackage{marginnote}
\usepackage{xcolor}

\usepackage{etoolbox}
\newbool{fullpaper}
\booltrue{fullpaper}

\usepackage[activate={true,nocompatibility},final,tracking=true,kerning=true,spacing=true,stretch=10,shrink=30]{microtype}
\microtypecontext{spacing=nonfrench}

\usepackage{balance}

\usepackage{subcaption}

\usepackage[nocompress]{cite}
\usepackage[hidelinks]{hyperref}

\usepackage{listings}
\usepackage{mathtools} 

\usepackage{proba}
\newcommand{\probability}[1]{\prob{[#1]}}

\newtheorem{lemma}{Lemma}
\newtheorem{theorem}{Theorem}

\newcommand{\todo}[1]{ \textbf{TODO:}\textit{#1} }
\renewcommand{\todo}[1]{}

\newcommand{\Oh}[1]{\ensuremath{\mathcal{O}(#1)}}
\newcommand{\OhL}[1]{\ensuremath{\mathcal{O}\!\left(#1\right)}}

\newcommand{\set}[2]{\{#1,\ldots,#2\}}
\newcommand{\myset}[2]{#1..#2}

\begin{document}

\title{Lightweight MPI Communicators with Applications to Perfectly Balanced Quicksort}

\author{\IEEEauthorblockN{Michael Axtmann}
\IEEEauthorblockA{Karlsruhe Institute of Technology\\
Karlsruhe, Germany\\
michael.axtmann@kit.edu}
\and
\IEEEauthorblockN{Armin Wiebigke}
\IEEEauthorblockA{Karlsruhe Institute of Technology\\
Karlsruhe, Germany\\
arminwiebigke@gmail.com}
\and
\IEEEauthorblockN{Peter Sanders}
\IEEEauthorblockA{Karlsruhe Institute of Technology\\
Karlsruhe, Germany\\
sanders@kit.edu}
}

\maketitle

\begin{abstract}
  MPI uses the concept of communicators to connect groups of processes.
  It provides nonblocking collective operations on communicators to overlap communication and computation.
  Flexible algorithms demand flexible communicators.
  E.g., a process can work on different subproblems within different process groups simultaneously, new process groups can be created, or the members of a process group can change.
  Depending on the number of communicators, the time for communicator creation can drastically increase the running time of the algorithm.
  Furthermore, a new communicator synchronizes all processes as communicator creation routines are blocking collective operations.
  
  We present RBC, a communication library based on MPI, that creates range-based communicators in constant time without communication.
  These RBC communicators support (non)blocking point-to-point communication as well as (non)blocking collective operations.
  Our experiments show that the library reduces the time to create a new communicator by a factor of more than 400 whereas the running time of collective operations remains about the same.
  We propose Janus Quicksort, a distributed sorting algorithm that avoids any load imbalances.
  We improved the performance of this algorithm by a factor of 15 for moderate inputs by using RBC communicators.
  Finally, we discuss different approaches to bring nonblocking (local) communicator creation of lightweight (range-based) communicators into MPI.
\end{abstract}

\section{Introduction}

The size of supercomputers rapidly increased to petascale machines with millions of cores.
The de facto standard for communication in High Performance Computing (HPC) is the Message Passing Interface (MPI).
Many applications need a flexible management of process groups, e.g., to adjust the scope of parallelism for load balancing, to achieve parallelism on multiple levels, or to divide tasks into fine-grained subproblems~\cite{Dinan2011}~\cite{balaji2009mpi}.
MPI uses the concept of communicators to enable multiple levels of parallelism by connecting groups of processes.
MPI provides (collective) communication operations between processes of a communicator to ensure scalability, portability, and comfortable programming with a high-level interface~\cite{hoefler2007implementation,bala1995ccl}.
The group context of a communicator guarantees that collective communication and point-to-point communication within one communicator as well as over different communicators does not interfere.
Since the introduction of MPI standard version 3.0 (MPI-3.0), communicators support both, nonblocking collective operations, and nonblocking point-to-point communication.
However, creating a new communicator still remains a blocking collective operation over the processes in the new group.

Collective communicator creation has multiple disadvantages.
First, as a blocking collective operation, it synchronizes all processes of the new process group.
This can cause immense idle time if a process starts the communicator creation behind schedule.
Moreover, as each process must create one communicator after another, worst case construction sequences can delay communicator creation or cause deadlocks if communicators overlap.
Thus, the processes must agree on a schedule to create new communicators.
Second, the most recent open-source implementations Open MPI 3.0 and MPICH-3.2 create an array of process IDs when the user creates a communicator.
In this case, the construction time of a communicator is linear to the number of group members.
Mohamad Chaarawi and Edgar Gabriel~\cite{chaarawi2008evaluating} integrated sparse data storages into Open MPI.
Their implementation reduces the footprint of an existing communicator. But the process group is stored explicitly during the communicator construction.
Finally, MPI does not provide a method to invoke nonblocking collective operations on a subset of processes without creating a new communicator.
Before performing the collective operation, the user must create a communicator of the subset of processes with a blocking communicator creation routine.

The main contribution of this paper is the lightweight library \emph{RangeBasedComm} (RBC) based on MPI.
RBC creates new communicators, containing a process range of a parent communicator, in constant time without communication.
Our range-based communicators provide (non)blocking collective operations and point-to-point communication.
As RBC can not access the context ID of a message, the library does not fully support the nonblocking model of the MPI-standard.
Even though we restricted the semantics of communication, the library is already applicable to many algorithms.
Furthermore, we discuss possible extensions of MPI implementations to provide nonblocking communicator creation without those restrictions.
We performed a large number of experiments on thousands of processes which show the advantage of lightweight communicators.
Our experiments show that our library reduces the time to create a new communicator by multiple orders of magnitude whereas the running time of collective operations remains about the same.

We furthermore propose, implement, and evaluate a new load balancing approach for recursive algorithms.
A straightforward approach of multi-level algorithms is to assign subtasks to disjoint subsets of processes.
As a result, the algorithms may apply expensive mechanisms in advance to reduce load imbalances between process groups.
We propose the approach of \emph{janus processes} for multi-level algorithms.
These janus processes work on two subtasks to minimize load imbalances.
Both subtasks are processed simultaneously to avoid that progress in one subtask delays progress in another subtask.
Our new sorting algorithm, \emph{Janus Quicksort} (JQuick), applies this approach to distributed sorting algorithms.
In comparison to hypercube quicksort~\cite{wagar1987hyperquicksort}, JQuick runs on any number of cores, rather than being restricted to a power of two and avoids any data imbalances.
Our experiments show that lightweight local communicator creation with RBC, instead of blocking communicator creation with native MPI, speeds up JQuick by multiple orders of magnitude for moderate input sizes.

\emph{Paper Overview.} We give preliminary definitions in Section~\ref{s:prelim}.
Then, we discuss MPI communicators and collective operations (Section~\ref{s:comms and colls}) as well as load balancing problems of massively parallel sorting algorithms (Section~\ref{s:load balancing probs}).
Section~\ref{s:rbc lib} describes our RBC library in detail.
We propose a nonblocking communicator creation function for the MPI standard in Section~\ref{s:range comms in mpi}. We also give an implementation recommendation that even runs local in constant time if the new process group is a range of processes of the parent communicator.
Our new perfectly balanced quicksort algorithm is described and analyzed in Section~\ref{s:squick}.
Section~\ref{s:experiments} gives an extensive experimental evaluation.

\section{Preliminaries}\label{s:prelim}

The input of sorting algorithms are $n$ elements on $p$ processes with $\Oh{n/p}$ elements each.
The output must be globally sorted, i.e., each process has elements with consecutive ranks.
We also want $\Oh{n/p}$ output elements on each process.

A common abstraction of communication in supercomputers is the (symmetric) single-ported message passing model.
It takes time $\alpha + l\beta$ to send a message of size $l$ machine words.
The parameter $\alpha$ defines the startup overhead of the communication.
The parameter $\beta$ defines the  time to communicate a machine word.
For simplicity, we assume that the size of a machine word is equivalent to the size of a data element.
For example, broadcast, reduction, and prefix sums can be implemented to run in time $\Oh{\beta l + \alpha \log p}$~\cite{bala1995ccl, sanders2009two} for vectors of size $l$.
We have $\alpha\gg\beta\gg 1$ where our unit is the time for executing a simple machine instruction.
Most of the time, we treat $\alpha$ and $\beta$ as variables in our asymptotic analysis in order to expose effects of latency and bandwidth limitations.

For simplicity, we will assume that all elements have unique keys.
This is without loss of generality in the sense that we can enforce this assumption by an appropriate tie-breaking scheme.
\iffullpaper
For example, replace a key $x$ with a tuple $(x, y)$ where $y$ is the global position in the input array.
With some care, this can be implemented with negligible overhead in such a way that $y$ does not have to be stored or communicated explicitly~\cite{wiebigke2017schizo}.
\fi

\section{Communicators and Collective Operations}\label{s:comms and colls}

Collective operations and point-to-point communication are executed in the context of a specific communicator.
A communicator stores an unique context ID and a group of $p$ processes with ranks $\myset{1}{p}$.\footnote{We use the notation $\myset{a}{b}$ as a shorthand for $\set{a}{b}$.}
The context ID guarantees that communication over one communicator as well as over different communicators does not interfere.
When a process sends a message over a communicator $c$, MPI stores the context ID of that communicator in the message header.
The receiver then matches the context ID of incoming messages with the context ID of $c$ to distinguish the expected message from messages send in a different context.
The context ID is managed by MPI and not visible to the user.

The user can create a new communicator based on a process subset of a parent communicator.
When a new communicator is created, the processes must agree on a new context ID that is not used by any process of that subset.
There exist different methods to create a new context ID~\cite{bala1995ccl, mpichdesigndoc}.
Open MPI and MPICH-3 use the concept of context ID masks.
A context ID mask is a bit vector which is used to track used context IDs.
Each process holds an own mask and the masks vary between processes depending on their communicators.
To find an unused context ID, the processes of the new group invoke a collective all-reduce operation on their context ID masks with \texttt{MPI\_BAND}.
Then the processes select the process ID which is represented by the least significant non-zero bit of the reduced bit vector.
According to the MPI standard~\cite{mpi30standard} it is theoretically possible to agree on a new context ID without communication for the case that each process of the parent communicator participates in the communicator creation.
This approach would require a larger context name space.
Furthermore, it is unknown whether this statement holds if just the processes of the new communicator invoke the communicator creation routine.

MPI provides two methods to create a communicator based on a process subset of a parent communicator.
In the first method, the user invokes the operations \texttt{MPI\_Comm\_create} or \texttt{MPI\_Comm\_create\_group}. 
The operation \texttt{MPI\_Comm\_create} is a blocking collective operation on the parent communicator whereas \texttt{MPI\_Comm\_create\_group} is a blocking collective operation on the processes of the new communicator~\cite{Dinan2011}.
As a parameter, those operations expect an MPI group that stores the new group of processes.
The user has two possibilities to create an MPI group -- either by enumerating the ranks explicitly (\texttt{MPI\_Group\_incl}) or by providing a sparse representation of rank ranges (\texttt{MPI\_Group\_range\_incl}).

The second method is to invoke the operation \texttt{MPI\_Comm\_split}.
This operation must be called by all processes of the parent communicator.
Each process passes its communicator affiliation (color) and its new rank (key).
MPI groups the processes by color and creates one communicator each.
When \texttt{MPI\_Comm\_split} is invoked, the processes create the process group collectively.
Open MPI and MPICH perform an all-gather operation to route all colors and keys to each process.
Then each process uses this information to calculate its group locally.
The all-gather operation takes $\Omega(\alpha\log p + \beta p)$ time which does not scale for small process subgroups.
Thus, the user should only use the operation \texttt{MPI\_Comm\_split} if the processes do not have all information to generate an MPI group just locally.
Gropp and Sack~\cite{sack2010scalable} propose communicator construction by parallel sorting in $\Theta(\sqrt{p})$ time with the result that a single message exchange of $n$ elements takes $\Omega(\alpha\log p + \beta n)$ time in worst case.
Siebert and Wolf~\cite{siebert2011parallel} reduced the time to construct a communicator to $\Oh{\log^2 p}$.

MPI groups store the mapping from rank to process ID.
E.g., MPICH-3 stores the mapping for each process explicitly in an array.
Mohamad~Chaarawi and Edgar~Gabriel~\cite{chaarawi2008evaluating} implemented different storage formats into Open MPI to reduce the memory footprint of a communicator.
E.g., the \emph{Range Format} derives from the syntax of \texttt{MPI\_Group\_range\_incl} and has constant access time and space (see also~\cite{kamal2010scalability}).
They always choose the format which minimizes memory consumption.
Unfortunately, their implementation still creates an explicit mapping with $\Oh{n}$ space during construction time.
In result, implementations of algorithms with polylogarithmic running time which split communicators are generally not possible using these libraries.

MPI-3.0 extends MPI by nonblocking collective operations~\cite{hoefler2006non}.
Torsten Hoefler and Andrew Lumsdaine~\cite{hoefler2007implementation} proposed a scheme to implement nonblocking collective operations with point-to-point communication.
They create a round-based schedule in which a round executes the collective operation until a data dependency prevents further computation.
The next round starts as soon as the data dependency is solved.
An additional communicator avoids tag conflicts between nonblocking collective operations and point-to-point communication that the user invoked.
A tag counter avoids tag conflicts between nonblocking collective operations.
The tag counter must be synchronous on all processes -- nonblocking collective operations on a subset of the processes would unsynchronize the tag counter.

\section{Massively Parallel Sorting}\label{s:load balancing probs}

For sorting large inputs, there are algorithms which move the data only once. Parallel sample sort~\cite{blelloch1991comparison} is a generalization of quicksort to $p-1$ pivots which are chosen from a sufficiently large sample of the input. Each process partitions its local data into $p$ pieces using the pivots and sends piece $i$ to process $i$.
After the all-to-all exchange, each process sorts its received pieces locally.
Since every process receives $p-1$ pivots, sample sort can only be efficient for $n=\Omega(p^2 / \log p)$ (see~\cite{kumar1994introduction}).
Indeed, the involved constant factors can be fairly large since the all-to-all exchange implies $p-1$ message startups if data exchange is done directly.

Algorithms with polylogarithmic running time are practical for small $n/p$.
Hypercube quicksort~\cite{wagar1987hyperquicksort} is a recursive algorithm on $2^k$ processes which performs $k$ levels of recursion.
On each level, the processes agree on a pivot and partition their data into two pieces according to the pivot.
Then, the pieces with small elements are routed to processes $\myset{1}{p/2}$ and pieces with large elements are routed to processes $\myset{p/2}{p}$.
Finally, hypercube quicksort is executed on the left and the right group of processes.

Compromises between these two extremes -- high asymptotic
scalability but a logarithmic number of data exchanges versus low
scalability but only a single communication -- have been considered.
E.g.,  multi-level variants of sample sort~\cite{gerbessiotis1994direct} agree on $k-1$ pivots,
partition local data into $k$ pieces, route piece $i$ to process group $i$
and recursively invoke sample sort on each process group.

The algorithms described above have several disadvantages.
First, they partition data into buckets and assign those buckets to process groups of fixed size.
In result, the workload between the process groups is not balanced or the workload is balanced but comes at the price of expensive pivot selection.
Second, the user can not execute the algorithms on arbitrary numbers of processes as the algorithms partition the processes into subgroups of equal size.
Finally, recursive implementations~\cite{sundar2013hyksort,axtmann2015practical,axtmann2017robust} create new communicators on each level for the sake of simplicity.
This usually prohibits polylogarithmic running time.

\section{Nonblocking Communication on Process Ranges}\label{s:rbc lib}

\begin{table}[t]
  \caption{Operation names and class names of RBC.}
  \label{tb:operations}
  \centering
  \begin{tabular}{|l|l|l|}
    \hline
    \textbf{Blocking Ops} & \textbf{Nonblocking Ops}         & \textbf{Classes} \\ \hline
    \texttt{rbc::Bcast}        & \texttt{rbc::Ibcast}                  & \texttt{rbc::Request} \\ \hline
    \texttt{rbc::Reduce}       & \texttt{rbc::Ireduce}                 & \texttt{rbc::Comm}    \\ \hline
    \texttt{rbc::Scan}         & \texttt{rbc::Iscan}                   &                  \\ \hline
    \texttt{rbc::Gather}       & \texttt{rbc::Igather}                 &                  \\
    \texttt{rbc::Gatherv}      & \texttt{rbc::Igatherv}                &                  \\ \hline
    \texttt{rbc::Barrier}      & \texttt{rbc::Ibarrier}                &                  \\ \hline
    \texttt{rbc::Send}         & \texttt{rbc::Isend}                   &                  \\
    \texttt{rbc::Recv}         & \texttt{rbc::Irecv}                   &                  \\ \hline
    \texttt{rbc::Probe}        & \texttt{rbc::Iprobe}                  &                  \\
    \texttt{rbc::Wait}         & \texttt{rbc::Test}                    &                  \\
    \texttt{rbc::Waitall}      &                                  &                  \\ \hline
                          & \texttt{rbc::Create\_RBC\_Comm} &                  \\
                          & \texttt{rbc::Split\_RBC\_Comm} &                  \\
                          & \texttt{rbc::Comm\_rank}              &                  \\
                          & \texttt{rbc::Comm\_size}              &                  \\ \hline
  \end{tabular}
\end{table}

We present the library \emph{RangeBasedComm} (RBC) based on MPI.
The key feature of the library is that \emph{RBC communicators} are created in constant time without communication.
A RBC communicator $R$ is derived from an MPI communicator $M$ and includes processes with ranks $\myset{f}{l}$ in $M$.
RBC provides (non)blocking point-to-point communication operations and (non)blocking collective operations in the context of an RBC communicator.
Table~\ref{tb:operations} gives a list of supported operations, methods to create RBC communicators, and classes of the RBC library.
We show example code how to use our library in \autoref{fg:sample}.
Because the interface is identical to MPI for most operations, our library can easily be integrated to replace existing MPI code.
Unless stated otherwise, the interface of our operations is identical to the interface of its equivalent in MPI.
The library is implemented in the namespace \texttt{rbc}.

\begin{figure}[t]
  \begin{lstlisting}
 int root = 0, e = 0, f, l, r, s;
 rbc::Comm world, range;
 rbc::Create_RBC_Comm(MPI_COMM_WORLD, &world);
 rbc::Comm_rank(world, &r);
 rbc::Comm_size(world, &s);
 if (r < s / 2) {f = 0; l = s / 2 - 1;}
 else {f = s / 2; l = s - 1;}
 // Local op. No synchronization.
 rbc::Split_RBC_Comm(world, f, l, &range);
 rbc::Request req; int flag;
 rbc::Ibcast(&e, 1, MPI_INT, r, range, &req);
 while (!flag) {
   // Do something else.
   rbc::Test(&req, &flag, MPI_STATUS_IGNORE);
 }
  \end{lstlisting}
  \caption{Nonblocking broadcast from rank $0$ to ranks ${\myset{0}{\text{s}/2 - 1}}$ and from rank $\text{s}/2$ to ranks $\myset{\text{s}/2}{\text{s} - 1}$. Both RBC communicators are created locally without process synchronization.}
  \label{fg:sample}	
\end{figure}

\subsection{RBC Communicator}\label{brc communicator}

A RBC communicator $R$ stores an MPI communicator $M$, the rank $f$ of its first process in $M$, and the rank $l$ of its last process in $M$.
The size of $R$ is defined by $(l - f) + 1$, the number of processes contained in $R$.
The \emph{RBC rank} of a process in $R$ with \emph{MPI rank} $m$ in $M$ is defined by $m - f$ and vice versa.
Note that more complicated projections from MPI rank to RBC rank are possible, e.g., a strided range\footnote{We implemented strided ranges but explain continuous ranges here.} with a stride factor of $s$ would contain MPI ranks $\myset{f,f+s}{f + s\lfloor (l-f)/ s\rfloor}$.
The local operation
\begin{lstlisting}
  rbc::Create_RBC_Comm(MPI_Comm&, rbc::Comm&)
\end{lstlisting}
creates an RBC communicator that contains all processes of an MPI communicator.
The local operation
\begin{lstlisting}
  rbc::Split_RBC_Comm(rbc::Comm&, rbc::Comm&,
                             int f, int l)
\end{lstlisting}
creates a new RBC communicator that contains processes with ranks $\myset{f}{l}$ of an existing RBC communicator.

As RBC can not access the context ID of an MPI message, the library weakens the semantics of RBC communicators.
We restrict the usage of tags if two RBC communicators overlap on more than one process, meaning multiple processes are part of both communicators.
All simultaneously executed communication operations have to use unique tags.
This is required to distinguish between operations in different RBC communicators which base on the same MPI communicator. 
If at most one process is part of both communicators, the communication on both communicators does not interfere.
In this case we do not restrict the usage of tags.

\subsection{Nonblocking Communication Requests}

Each nonblocking communication call returns a request of type \emph{\texttt{rbc::Request}}.
\texttt{rbc::Request} is a smart pointer to a request that implements the specific nonblocking operation.
When we talk about an object of type \texttt{rbc::Request} in the following, we mean the request object of the specific nonblocking operation.
When we invoke a nonblocking communication operation, we do not guarantee that the communication is completed once the operation returns.
Instead, the user has to invoke the method \texttt{rbc::Test} on the request to check if the operation is completed or not.
Our library also provides three additional methods to test nonblocking operations for completion.
The operation \texttt{rbc::Wait} takes a request and repeatedly calls \texttt{rbc::Test} until the operation is completed.
The operation \texttt{rbc::Testall} takes an array of requests, calls \texttt{rbc::Test} on all requests, and returns true if all operations are completed.
The operation \texttt{rbc::Waitall} takes an array of requests and repeatedly calls \texttt{rbc::Testall} until all operations are completed.

\subsection{Point-to-point Communication}

\begin{figure}
  \begin{lstlisting}
 rbc::Send(const void *sendbuf, int count,
     MPI_Datatype, int r, int t, rbc::Comm)
 int rbc::Isend(const void *sendbuf, int count,
     MPI_Datatype, int r, int t, rbc::Comm,
     rbc::Request *);
 rbc::Recv(void *buffer, int count,
     MPI_Datatype, int r, int t, rbc::Comm);
 int rbc::Irecv(void *buffer, int count,
     MPI_Datatype, int r, int t,rbc::Comm,
     rbc::Request *);
 rbc::Probe(int r, int t, rbc::Comm,
     MPI_Status *);
 int int rbc::Iprobe(int r, int t, rbc::Comm,
     int *flag, MPI_Status *);
 int rbc::Test(rbc::Request *request,
     int *flag, MPI_Status *status);
  \end{lstlisting}
  \caption{Interface of RBC for the operations used in point-to-point communication with rank \texttt{r} and tag \texttt{t}.}
  \label{fg:point-to-point}
\end{figure}

RBC provides (non)blocking send and receive operations.
\autoref{fg:point-to-point} gives the interface of these operations.
RBC point-to-point communication routines call MPI routines internally.
When the user invokes RBC with a specific tag and an RBC target rank, we call MPI with the corresponding MPI rank and the tag passed by the user.

\subsubsection*{Probing}

The operation \texttt{rbc::Iprobe} is a nonblocking operation that tests whether a message from a source process is ready to be received.
If the user invokes \texttt{rbc::Iprobe} with a specific rank, we invoke the operation \texttt{MPI\_Iprobe} with the parameters provided by the user and return when \texttt{MPI\_Iprobe} returns.
If the wildcard  \texttt{MPI\_ANY\_SOURCE} is used instead of a specific rank, we also invoke the operation \texttt{MPI\_Iprobe}.
However, if \texttt{MPI\_Iprobe} returns true, we only know that any message is ready to be received.
In this case, we test if the source process of that message is part of the RBC communicator.
We return true if the source process is in the RBC communicator, otherwise we return \texttt{false}.

The operation \texttt{rbc::Probe} waits until a message is ready to be received.
If the user calls this operation with a specific rank, we invoke the operation \texttt{MPI\_Probe}.
If the wildcard  \texttt{MPI\_ANY\_SOURCE} is used instead of a specific rank, we repeatedly call the operation \texttt{rbc::Iprobe} with our own input parameters until the operation returns \texttt{true}.
In both cases, we return the status that was returned by the operation  \texttt{MPI\_Probe} (specific rank) or \texttt{rbc::Iprobe} (wildcard).

Our implementation of \texttt{rbc::Probe} and \texttt{rbc::Iprobe} on wildcards is not straightforward.
In return, our library guarantees that communication over two different RBC communicators of the same MPI communicator does not interfere if the communicators overlap on at most one process.

\subsubsection*{Sending}

The operation \texttt{rbc::Send} is a blocking operation that sends a message to a destination process.
RBC forwards the input arguments of the operation \texttt{rbc::Send} to the operation \texttt{MPI\_Send}.
The operation \texttt{rbc::Isend} is a nonblocking send operation.
RBC internally calls the operation \texttt{MPI\_Isend} with the input arguments of \texttt{rbc::Isend} but replaces the RBC request, passed by the user, with an MPI request.
After the MPI operation returned, RBC stores the MPI request in the RBC request.
When the user invokes \texttt{rbc::Test} with the RBC request, we internally calls \texttt{MPI\_Test} on the stored MPI request.

\subsubsection*{Receiving}

The operation \texttt{rbc::Recv} is a blocking operation that receives a message from a source process.
The user can either call \texttt{rbc::Recv} with a specific source rank or with \texttt{MPI\_ANY\_SOURCE}.
If the wildcard is used, we firstly need to wait for an incoming message sent over the same RBC communicator and determine its source rank.
For this, we invoke the operation \texttt{rbc::Probe} which returns a status containing the source rank.
In both cases, we invoke the operation \texttt{MPI\_Recv} with the source rank.

The operation \texttt{rbc::Irecv} is a nonblocking receive operation.
The user can either call \texttt{rbc::Irecv} with a source rank or with \texttt{MPI\_ANY\_SOURCE}.
If \texttt{rbc::Irecv} is invoked with a specific source rank, we call the operation \texttt{MPI\_Irecv} with the input arguments of \texttt{rbc::Irecv} but replace the RBC request, passed by the user, with an MPI request.
When the operation returns, we store the MPI request in the RBC request.
When the user invokes \texttt{rbc::Test} with the RBC request, we internally call \texttt{MPI\_Test} on that MPI request.
If a wildcard is used instead of a specific rank, we search for a message which is sent over the same RBC communicator by invoking the operation \texttt{rbc::Iprobe}.
If \texttt{rbc::Iprobe} returns true, we know that a message is ready to be received and call \texttt{rbc::Irecv} with the source rank of that message.
Afterwards, we return the RBC request created by the operation \texttt{rbc::Irecv}.
Otherwise, there is no message that is ready to be received.
We return a request that again searches for incoming messages when the operation \texttt{rbc::Test} is invoked with that request.
Once an incoming message is detected, we call \texttt{rbc::Irecv} with the source rank of that message.
Afterwards, we return the RBC request created by the operation \texttt{rbc::Irecv}.

\subsection{Collective Operations}

The communication of collective operations involves all processes of the RBC communicator.
This means that all processes have to call the operation or the collective operation will not be executed completely.
Collective operations are implemented with point-to-point communication provided by the RBC library.
All processes of the same communicator must call all (non)blocking collective operations in the same order.
All implementations exploit binomial tree based communication patterns~\cite{vadhiyar2000automatically, blelloch1990prefix}.
The communication patterns are generic, not optimized for a specific network, but theoretically optimal for small input sizes~\cite{chan2007collective}.
It is easy to extend our library by additional collective operations, e.g., for large input sizes.

When a process has completed a blocking collective operation locally, the process can invoke a new blocking collective operation.
Likewise, when a process has completed a nonblocking collective operation locally (\texttt{rbc::Test} returned true), the process can invoke a new nonblocking collective operation.
This rule even holds when outgoing messages of the completed collection are still pending, i.e., a message transfer of the collective operation has been invoked asynchronously and the message content is still stored in a send buffer of MPI.
The MPI standard guarantees that two messages, sent from process $i$ to process $j$, are received by process $j$ in the send order.

\subsubsection*{Blocking collectives}

We define a distinct exclusive tag for each blocking collective operation.
RBC uses this tag internally to perform point-to-point operations.
As long as the user does not use these reserved tags, blocking collective operations do not interfere with other communication.

\subsubsection*{Nonblocking Collectives}

We define a distinct exclusive tag for each nonblocking collective operation.
Alternatively, the user can specify an own user-defined tag, e.g.,
\begin{lstlisting}
  int rbc::Ibcast(void *buffer, int count,
      MPI_Datatype, int root, rbc::Comm,
      MPI_Request *, int tag = RBC_IBCAST_TAG);
\end{lstlisting}
This avoids interferences between simultaneously executed nonblocking collective operations on the same RBC communicator as well as between different RBC communicators.
Note that a preserved tag-space, as used in many MPI implementations, would not avoid interference of nonblocking operations between different RBC communicators if they base on the same MPI communicator and share processes.

We implement our nonblocking collective operations with state machines, similar to the approach in~\cite{hoefler2007implementation}.
Each state begins with local work, e.g., applying the reduce operator, and ends with pending send/receive operations if these operations introduce a data dependency.
A send operation causes a data dependency if the send buffer is required for further computations.
A receive operation causes a data dependency if subsequent operations rely on data of different processes.
When the user invokes a nonblocking collective operation,  RBC creates a request object which contains a local state machine, executes its first state, and returns the request.
To make further progress on that operation, the user has to call the operation \texttt{rbc::Test} (or one of the three other test operations) on the corresponding request.
This operation checks whether outstanding data dependencies remain.
If so, we can not proceed and we return \texttt{false}.
Otherwise, we return \texttt{true} if the collective operation is already finished locally or we execute the next state and return \texttt{false} afterwards.

\section{Nonblocking Creation of (Range-Based) MPI Communicators}\label{s:range comms in mpi}

\begin{figure*}[!t]
\centering
\includegraphics{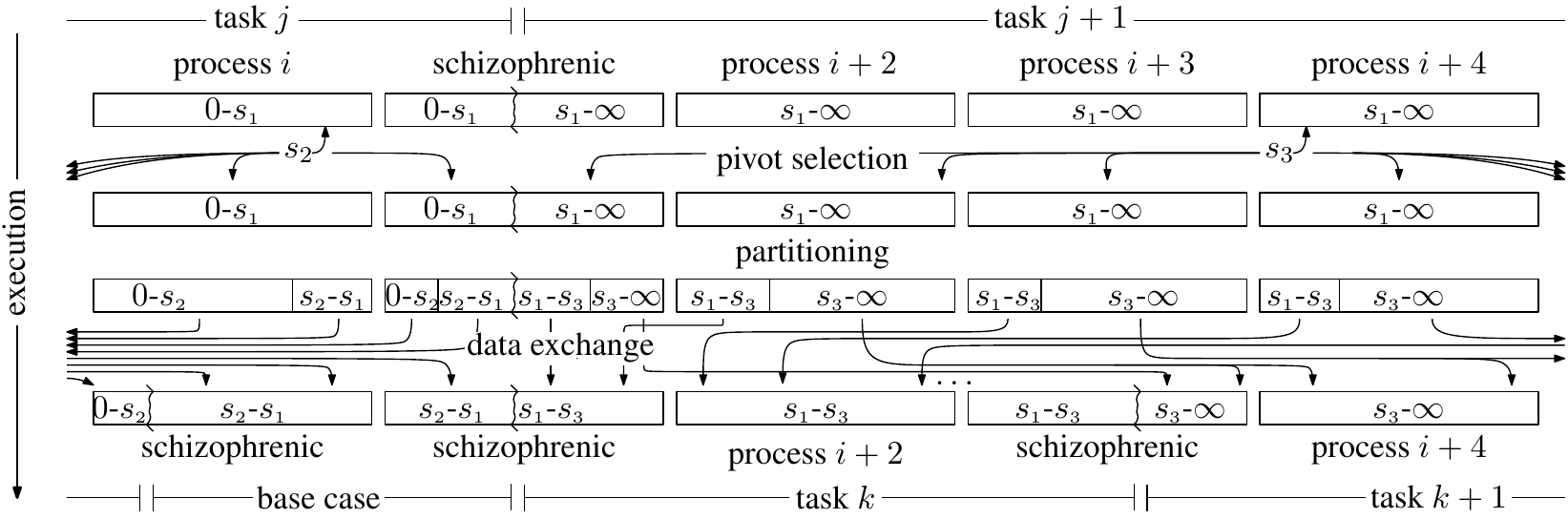}
\caption{Algorithm scheme of JQuick.}
\label{f:squick algo scheme}
\end{figure*}

RBC weakens the semantics of RBC communicators as it can not access the context ID of an MPI message (see Section~\ref{brc communicator}).
In this section we recommend the function
\begin{minipage}{\linewidth}
\begin{lstlisting}
  int MPI_Icomm_create_group(
    MPI_Comm, MPI_Group, int tag,
    MPI_Comm *, MPI_Request *)
\end{lstlisting}
\end{minipage}
as a part of the Message Passing Interface and give an implementation proposal which does not weaken the semantics of MPI.
Similar to the function \texttt{MPI\_Comm\_create\_group} of MPI, the new function is collectively invoked by the process group of the new communicator.
But it is a nonblocking function which can be implemented as a local operation that runs in constant time if the new process group is a range of processes of the parent communicator.
We recommend that the user creates an MPI group by invoking \texttt{MPI\_Group\_range\_incl} if appropriate.
If the new MPI group contains only few ranges, \texttt{MPI\_Group\_range\_incl} shall use the Range Format proposed by Mohamad Chaarawi and Edgar Gabriel~\cite{chaarawi2008evaluating} to construct the group in time linear to the number of ranges.
Communicators constructed with our new function do not weaken the semantics of MPI communication operations. E.g. the communication of one MPI communicator does not influence communication of other MPI communicators.
\texttt{MPI\_Icomm\_create\_group} fully supports the nonblocking collective communication model.
This model requires all processes of one communicator to call nonblocking collective operations in the same order. 

Implementations of \texttt{MPI\_Icomm\_create\_group} boil down to agree on a distinct context ID not used by any process of the MPI group.
Our implementation proposal uses context IDs consisting of four integers $\langle a, b,f,l,c\rangle$ and each process maintains a counter, initialized to $0$.
When the user invokes \texttt{MPI\_Icomm\_create\_group}, a new context ID is calculated.
If the \texttt{MPI\_Group} parameter does not represent a continuous range of processes of its parent communicator, the first process of \texttt{MPI\_Group} calculates a new context ID $\langle a, b, 0, l, 0\rangle$ where $a$ is its process ID, $b$ the value of its counter, and $l$ the number of processes in \texttt{MPI\_Group}.
Then the first process increases its counter and performs a nonblocking broadcast operation with the user-defined tag to send the new context ID to the remaining processes.
This operation runs in $\Oh{\alpha \log l}$ time.
Otherwise, \texttt{MPI\_Group} does represent a continuous range of ranks $\set{f'}{l'}$ of its parent communicator with the context ID $\langle a, b,f,l,c\rangle$.
In this case, the new context ID $\langle a, b, f + f', f + l',c+1\rangle$ is calculated locally in constant time on each process.
If $f'=0$ and $l' = l-f$, the new communicator and its parent communicator contain the same group of processes and the value $c+1$ is used to distinguish their context IDs.

If the user creates two communicators simultaneously on the same parent communicator, both operations make progress simultaneously as MPI overlaps both broadcast operations.
A nonblocking version of \texttt{MPI\_Comm\_create} can be implemented as described above but the nonblocking broadcast operation is invoked by all processes of the parent communicator and thus does not need a user-defined tag.
Note that context IDs only require the first two integers of the context ID described above if the nonblocking communicator creation does not take advantage of the range-based case.
We do not recommend to calculate a new context ID by performing a nonblocking all-reduce operation on context ID masks.
In this case, a process can not make progress on multiple nonblocking communicator creations simultaneously as the all-reduce operations of different nonblocking communicator creation routines must be executed one after another.

\section{Janus Quicksort}\label{s:squick}

We present \emph{Janus Quicksort} (JQuick),  a recursive sorting algorithm for distributed memory systems which is based on Quicksort.
JQuick guarantees perfect data balance, i.e., after each level each process stores $\lfloor n/p\rfloor$ or $\lceil n/p\rceil$ elements.
Firstly, JQuick performs a distributed phase which recursively partitions tasks into subtasks.
Then a second phase sorts \emph{base cases}.
Base cases are subtasks covering only one or two processes.
Figure~\ref{f:squick algo scheme} depicts one distributed level of recursion.
We prove that the distributed phase of JQuick takes $\Oh{\alpha \log^2 p + \beta n/p \log p}$ time with probability $1-\Oh{p^{-6}}$ and that the base cases are executed in $\Oh{\alpha + \beta n/p + n/p\log (n/p)}$ time (Lemma~\ref{t:running time}).
For the sake of simplicity, we assume that $n$ is a multiple of $p$.

One level of recursion consists of four steps: pivot selection, data partitioning, data assignment, and data exchange.
In the first step, a random element is selected and broadcasted to all processes.
In the second step, the processes partition their local data into a \emph{left partition} of elements smaller than the pivot (small elements) and a \emph{right partition} of elements larger than or equal to the pivot (large elements).
In the third step, the processes assign elements to target processes.
Here, we describe a simple \emph{greedy message assignment} algorithm.
Firstly, a distributed exclusive prefix sum is performed over the number of small and large elements on each process.
Let the prefix sum on process $i$ be $s_i$, the number of small elements on processes $\myset{0}{i-1}$.
The number of large elements on processes  $\myset{0}{i-1}$ is $l_i = i\cdot n/p - s_i$.
The last process broadcasts $s_{p-1}$ and $l_{p-1}$, the total number of small and large elements.
Next, we split the processes into a \emph{left group} of processes $\myset{0}{\lfloor np/s_{p-1} \rfloor}$ and a \emph{right group} of processes $\myset{\lfloor np/s_{p-1}+1 \rfloor}{p-1}$.
Finally, we calculate a data assignment that assigns the small elements to the left group and the large elements to the right group.
Our assignment guarantees that each process gets exactly $n/p$ elements assigned.
The first source process assigns its small elements to target process $0$.
Then the second source process again assigns small elements to target process $0$ as long as this process has residual capacity left.
If the second source process still has small elements, it assigns the remaining small elements to the next target process.
We continue with the remaining source processes and also assign large elements in the same way to the right group.

Even though a process sends two messages to each group, a process may receive $\Theta(\min(p, n/p))$ messages in the worst case.
An alternative assignment algorithm, \emph{deterministic message assignment}~\cite{axtmann2017robust}, guarantees that each process sends at most eight messages and receives at most eight messages.
Both assignment algorithms take $\Oh{\alpha \log p}$ time.

Once we have calculated the data assignment we start the data exchange step.
Each process invokes a nonblocking send operation to its target processes and then receives messages until $n/p$ elements have been received.

After the elements have been redistributed, the left group recursively sorts the small elements and the right group recursively sorts the large elements.
If $\lfloor n/k \rfloor = \lfloor n/k+1 \rfloor$, the left and the right group are not disjoint.
In this case, process $\lfloor n/k \rfloor$ receives $s_{p-1} \mod n/p$ small elements and $n/p - s_{p-1} \mod n/p$ large elements.
We call this process a \emph{janus process}.
Our data assignment guarantees that the number of elements on a janus process in the left and the right group still sum up to $n/p$.
This janus process must proceed on both groups simultaneously.
Otherwise, progress in one group delays progress in another group, and so on.
Janus processes perform all local operations on both groups simultaneously before they communicate again.
All communication operations are then executed in nonblocking mode, again on both groups simultaneously.

After the distributed phase has been executed, the second phase starts sorting base cases.
This guarantees that a janus process does not delay the execution of a larger subtask while sorting a base case.
A base case on a single process is sorted locally.
When a base case is sorted by two processes, the left (right) process receives the elements from the right (left) process, performs the operation \emph{quickselect} to determine the small (large) elements according to its workload, and finally sorts those elements locally.

On recursion level $i>0$ the first and the last process of a process group might already be janus processes.
For those processes, the load of $n/p$ elements is distributed over a left group, a right group and potential base cases in the middle.
To calculate the data assignment correctly, each group of processes keeps track of the remaining load on the first process of its process group.
Let $r$ be the remaining load of the first process in a process group with $p$ processes.
The remaining load of the first process in the left subgroup remains $r$.
Each process in the right subgroup updates the remaining load of their first process to $r'=n/p - (n/p+s_{p-1}-r) \mod n/p$.

\subsection{Analysis}

The main result of the analysis are running time guarantees for the distributed phase and the base case phase of the algorithm.
We prove in Theorem~\ref{t:running time} that the distributed phase of JQuick takes $\Oh{\alpha \log^2 p + \beta n/p\log p}$ time with probability $1-\Oh{p^{-6}}$ and that the base cases take $\Oh{\alpha + \beta n/p + n/p\log (n/p)}$ time.
\iffullpaper
For our analysis we adapt the analysis from Jaja~\cite{jaja2000perspective} which shows that sequential quicksort with random pivot selection runs in $\Oh{n\log n}$ time with probability $1-\Oh{n^{-6}}$.
In a first step we show that JQuick performs $\Oh{\log p}$ levels of recursion until its data gets sorted locally with probability $1-\Oh{p^{-6}}$ (Lemma~\ref{l:rec depth}-\ref{l:total recursion depth}).

\begin{lemma}
  \label{l:rec depth}
  A randomly selected input element passes with probability $\Oh{p^{-7}}$ more than $20 \log p$ levels of recursion until it is part of a base case.
\end{lemma}

\begin{IEEEproof}
  We define a recursion step with global load $n$ as a \emph{successful recursion step} if both subtasks of JQuick have a load of at most $7/8n$ elements.
  Let $e$ be an arbitrary input element.
  Jaja~\cite{jaja2000perspective} has shown that random split of a sorting task with $n$ elements creates two subtasks of at most $7/8n$ elements each with probability $3/4$.
  From this follows that an arbitrary input element $e$ is part of a task with at most $l_e \leq (7/8)^kn$ elements after it passed $k$ successful levels of recursion.
  After $k = \log_{8/7} (p/2)$ successful recursion steps we have $l_e \leq (7/8)^kn = 2 n/p$ elements and execute the base case.

  We now show that a randomly selected input element passes at least $\log_{8/7} (p/2)$ successful recursion steps with probability $\Oh{p^{-7}}$ when $20 \log_{8/7} p$ recursion steps are executed.
  This random experiment is a Bernoulli trail as we have exactly two possible outcomes, ``successful recursion step'' and ``non-successful recursion step'', and the probability of success is the same on each level.
  Let denote the random variable $X$ as the number of non-successful recursion steps after $20 \log_{8/7} p$ recursions.
  The probability $I$
  \begin{equation}
    \begin{aligned}
      I &= \probability{X > 20 \log p - \log (p/2)}
      \leq \probability{X > 19 \log p}\\
      &\leq \sum_{j>19 \log p} \binom{20 \log p}{j}\left(\frac{1}{4}\right)^j\left(\frac{3}{4}\right)^{20 \log p - j}\\
      &\leq \sum_{j>19 \log p} \left(\frac{20 e \log p}{j}\right)^j\left(\frac{1}{4}\right)^j\\
      &\leq \sum_{j>19 \log p} \left(\frac{5 e \log p}{19 \log p}\right)^j
      = \sum_{j>19 \log p} \left(\frac{5 e}{19}\right)^j
      = \OhL{p^{-7}}
    \end{aligned}
  \end{equation}
  defines an upper bound of the probability that a randomly selected input element passes $20 \log_{8/7} p$ recursion steps without passing $\log_{8/7} (p/2)$ successful recursion levels.
  For the sake of simplicity, all logarithms of the equation above are to the base of $8/7$.
  The third ``$\leq$'' uses $\binom{n}{k} \leq \left(\frac{en}{k}\right)^k$ and the second ``$=$'' uses the geometric series.
\end{IEEEproof}

\begin{lemma}
  \label{l:total recursion depth}
  JQuick executes more than $\Oh{\log p}$ distributed levels of recursion with probability $\Oh{p^{-6}}$.
\end{lemma}

\begin{IEEEproof}
  Let $E = \myset{e_1}{e_{n}}$ be the input data of JQuick in sorted order and let $L = \{e_{in/p} | 1 i \leq p\wedge i \in \mathbb{N}\}$ be every $n/p$th element.
  Lemma~\ref{l:rec depth} and Boole's inequality imply that at least one element $e \in L$ does not run into the base case after $20 \log_{8/7} p$ recursions with a probability of $\Oh{p \cdot p^{-7}} = \Oh{p^{-6}}$.

  We analyze the length of the critical path in the recursion tree of JQuick.
  Let $g$ be the process group on the last level of recursion (before the base case) of that path.
  Process group $g$ contains at least $2 n/p$ elements.
  Those elements build a subsequence of $E$ and thus contain at least one element $e \in L$.
  Element $e$ participates on more than $20 \log_{8/7} p$ recursion steps with probability $\Oh{p^{-6}}$.
  Thus JQuick executes more than $20 \log_{8/7} p$ levels of recursion with probability $\Oh{p^{-6}}$.
\end{IEEEproof}
\else
\begin{lemma}
\label{l:total recursion depth}
JQuick executes $\Oh{\log p}$ distributed levels of recursion with probability $1-\Oh{p^{-6}}$.
\end{lemma}
\begin{IEEEproof}
  We only give the basic idea of a proof.
  We give an extended proof in the eponymous full paper~\cite{axtmann2017lightweight}.
  For our proof we adapt the analysis from Jaja~\cite{jaja2000perspective} which shows that sequential quicksort with random pivot selection runs in $\Oh{n\log n}$ time with probability $1-\Oh{n^{-6}}$.
  We call a distributed recursion step of our algorithm \emph{successful} if a random split of a task with $n$ elements creates two subtasks of at least $1/8n$ elements each.
  Obviously, a distributed recursion step is successful with probability $3/4$ and an element runs into the base case if it passes at least $\log_{8/7} (p/2)$ successful recursion steps.
  We use this observation in a Bernoulli trail to show that a randomly selected input element passes at least $\log_{8/7} (p/2)$ successful recursion steps with probability $\Oh{p^{-7}}$ when $20 \log_{8/7} p$ recursion steps are executed.
  Boole's inequality implies that at least one of every $n/p$th element does not run into the base case after $20 \log_{8/7} p$ recursions with a probability of $\Oh{p \cdot p^{-7}} = \Oh{p^{-6}}$.
  Now it is simple to prove that JQuick executes $\Oh{\log p}$ distributed levels of recursion with probability $1-\Oh{p^{-6}}$.
\end{IEEEproof}
\fi

\begin{theorem}
  \label{t:running time}
  For arbitrary inputs, the distributed phase of JQuick takes $\Oh{\alpha \log^2 p + \beta n/p\log p}$ time with probability $1-\Oh{p^{-6}}$ and the base cases take $\Oh{\alpha + \beta n/p + n/p\log (n/p)}$ time.
\end{theorem}

\begin{IEEEproof}
  We firstly analyze a single distributed recursion level on $p$ processes.
  The pivot selection step takes $\Oh{\alpha \log p}$ time, the data partitioning step takes $\Oh{n/p}$ time, and the data assignment step takes $\Oh{\alpha \log p}$ time.
  The data exchange step takes $\Oh{\alpha + \beta n/p}$ time if the deterministic message assignment algorithm is used.
  In this step each process sends (receives) a constant number of messages and sends (receives) exactly $n/p$ elements in total.
  Note that janus processes do not affect the asymptotic running time as communication of janus processes is always done simultaneously on both tasks with nonblocking operations.
  In total, a distributed recursion level takes $\Oh{\alpha \log p + \beta n/p}$ time.
  Lemma~\ref{l:total recursion depth} implies that all distributed levels of recursion take $\Oh{\alpha \log^2 p + \beta n/p\log p}$ time with probability $1-\Oh{p^{-6}}$.
  All base cases are sorted in $\Oh{\alpha + \beta n/p + n/p\log (n/p)}$ time.
  The base cases include a $\Oh{\alpha + \beta n/p}$ term as base cases are usually processed by two processes.
\end{IEEEproof}

\section{Experimental Results}\label{s:experiments}

We now present the results of our experiments.
We ran all experiments with our RBC library and with MPI implementations from Intel and IBM.
Our benchmarks have been executed on 32\,768 cores.
We divide our experiments into two sections.
In Section~\ref{sec:micro} we present microbenchmarks of RBC operations and the corresponding MPI operations.
Firstly, we present results for nonblocking collective operations executed with RBC and with its counterparts in MPI.
Then we compare the running time of creating communicators on non-overlapping and overlapping sub-ranges of processes with RBC and with native MPI.
Finally, we give results of collective operations executed on a sub-range of processes.
Section~\ref{sec:sorting benchmark} gives running times of JQuick executed with our RBC library and with native MPI.
We execute each microbenchmark five times.
We perform each experiment of JQuick seven times for $n/p \leq 2^{16}$ and three times for $n/p > 2^{16}$.
We report the average over all runs and use {$64$-bit} floating point elements.

We ran our experiments at the thin node cluster of the SuperMUC
(\href{www.lrz.de/supermuc}{www.lrz.de/supermuc}), an island-based distributed
system consisting of $18$ islands, each with 512 computation nodes. 
Each computation node has two Sandy Bridge-EP Intel Xeon E5-2680 8-core processors
with a standard frequency of $2.3$\,GHz
and $32$\,GByte of memory.
A non-blocking topology tree connects the nodes within an island
using the InfiniBand FDR10 network technology.
A pruned tree connects the islands
among each other with a bidirectional bisection bandwidth ratio of $4:1$.

\begin{figure}[t]
	\centering
	\includegraphics[width=\linewidth]{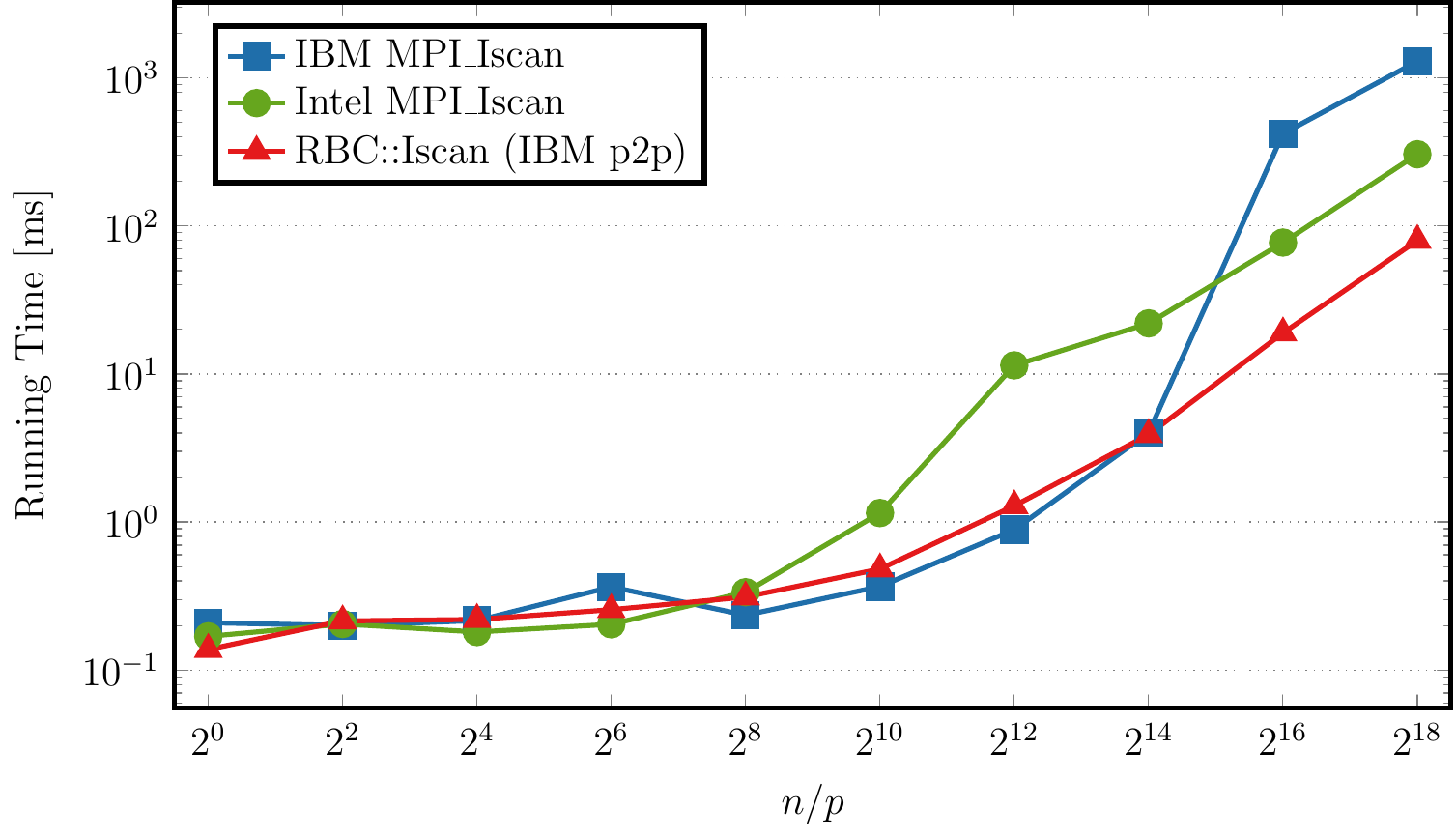}
	\caption{Running times of \texttt{MPI\_Iscan} and \texttt{rbc::Iscan} on $2^{15}$ cores with different number of input elements of type \texttt{double} on each process.}
        \label{fig:scan}
\end{figure}

\subsection{Implementation Details}

Our implementation of JQuick uses the greedy message assignment algorithm described above.
As a pivot we select the median of $\max(k_1\log p, k_2n/p, k_3)$ samples determined by the random sampling approach by Sanders et. al.~\cite{sanders2016efficient}.
We handle duplicates by carefully switching between the compare functions ``$<$'' and ``$\leq$'' as described in~\cite{wiebigke2017schizo}.
Janus Quicksort and the RBC library are written in C++ and compiled with version 16.0 of the Intel icpc compiler using the optimization flags \texttt{-O3 -ipo -xHost}.
For inter-process communication, we use either version 1.4 of the IBM MPI library or version 5.1.3 of the Intel MPI library.
Our initial experiments with IBM MPI have shown that the bulk transfer protocol increases fluctuations in running time.
We disable bulk transfer in the IBM MPI library by setting the environment variable \texttt{MP\_USE\_BULK\_XFER=no}.
Our implementation can be found at \url{https://github.com/MichaelAxtmann/RBC}.

\begin{figure}[t]
	\centering
	\includegraphics[width=\linewidth]{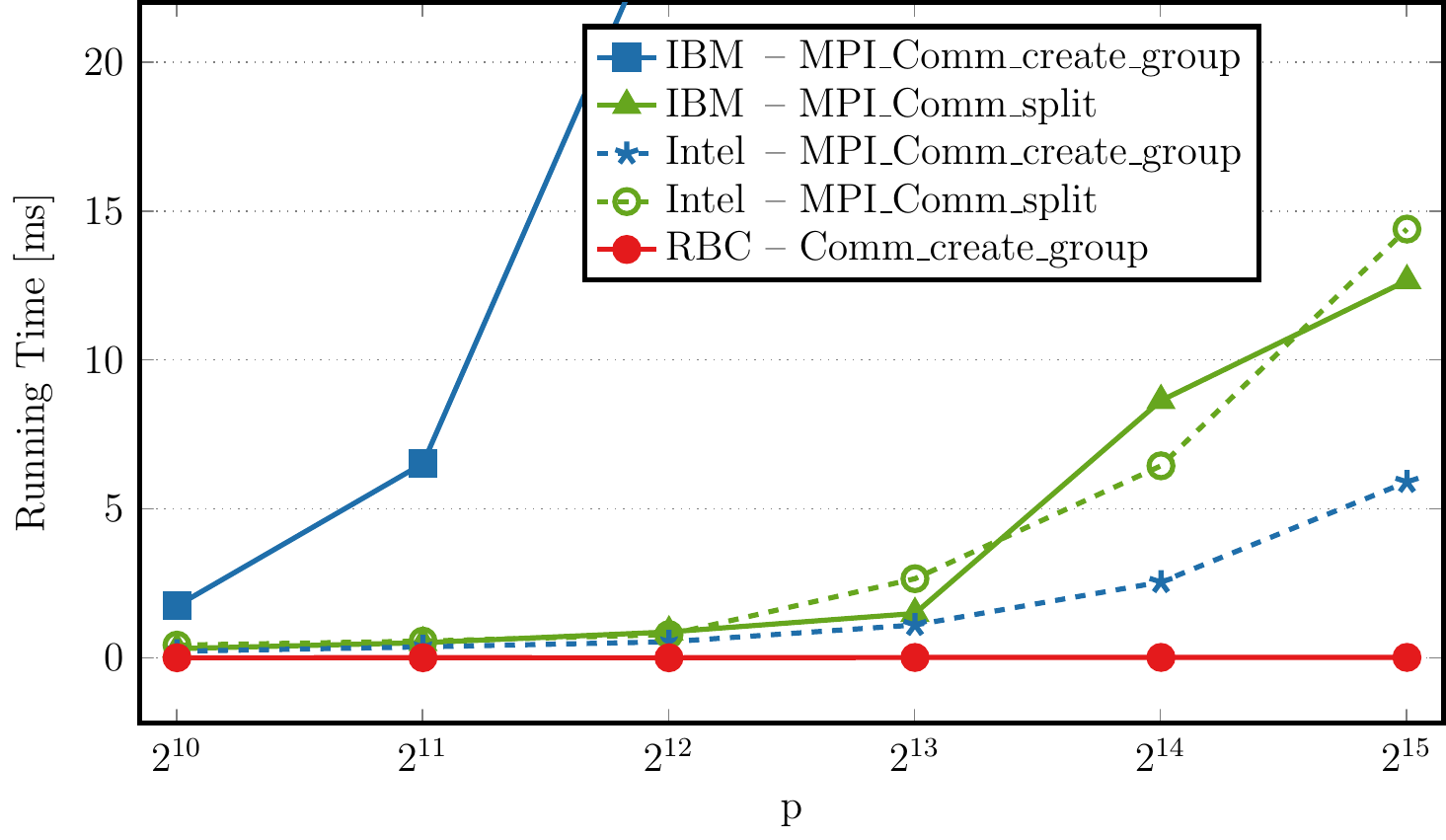}
	\caption{Running times of splitting a communicator of $p$ processes into one communicator containing processes $\myset{0}{p/2 - 1}$ and one communicator containing processes $\myset{p/2}{p-1}$.}
        \label{fig:split}
\end{figure}

\subsection{Microbenchmarks}\label{sec:micro}

\emph{Collective operations}. Figure~\ref{fig:scan} depicts the running times of the nonblocking operation \texttt{MPI\_Iscan} provided by native Intel MPI and native IBM MPI and its counterpart in the RBC library executed on $2^{15}$~cores.
For the sake of simplicity, we give the running times of the RBC library only on top of the IBM MPI as the running times on top of Intel MPI are almost the same.
We see that all implementations need about the same amount of time for moderate inputs, $n/p \leq 2^9$, where the running time is dominated by the message startup overhead.
For larger inputs, our library outperforms both MPI implementations by a factor of up to $16$.
\iffullpaper
We give running times of the collective operations broadcast, reduce, scan, and gather in Figure~\ref{fig:all colls} in the Appendix.
\else
We give running times of the collective operations broadcast, gather, and reduce in the eponymous full paper~\cite{axtmann2017lightweight}.
\fi
The experiments show that these operations executed with RBC perform similar to their counterparts in IBM MPI and Intel MPI.
In conclusion our range-based communicator creation does not come with hidden overheads in communication operations of RBC.

\emph{Communicator splitting}. Figure~\ref{fig:split} presents running times of splitting a parent communicator of $p$ processes into a communicator containing processes $\myset{0}{p/2 - 1}$ and a communicator containing processes $\myset{p/2}{p-1}$.
We perform this experiment with RBC communicators, MPI communicators created with \texttt{MPI\_Comm\_split}, and MPI communicators created with \texttt{MPI\_Comm\_create\_group}.
We invoke \texttt{MPI\_Comm\_create\_group} with an MPI group created with \texttt{MPI\_Group\_range\_incl}.
Note that the construction of an MPI communicator is a blocking collective function.
Furthermore, the function \texttt{MPI\_Comm\_split} must be invoked by all processes of the parent communicator.
In contrast to that we can split an RBC communicator in constant time without communication by calling the operation \texttt{rbc::Split\_RBC\_Comm} only using the processes of the new communicator.
Our experiments show that the time to construct an RBC communicator is negligible.
If MPI used a sparse representation of the process group, we could expect that the running time of \texttt{MPI\_Comm\_create\_group} increases logarithmically to the number of processes.
However, the running time of this operation executed with Intel MPI increases linearly with the number of processes.
This supports our hypothesis that Intel MPI represents MPI groups explicitly.
The running time of \texttt{MPI\_Comm\_create\_group} executed with IBM MPI is disproportionately slow and outperformed by the remaining operations by multiple orders of magnitude.
The implementation of \texttt{MPI\_Comm\_split} in Intel MPI and IBM MPI is slower than \texttt{MPI\_Comm\_create\_group} (Intel MPI) by a factor of two for large $p$.
We expected this slowdown as a process invokes \texttt{MPI\_Comm\_split} only with its own group affiliation.
Thus, this operation must construct an MPI group internally by gathering group information from each process.
          
\begin{figure}[t]
  \centering
  \includegraphics[width=\linewidth]{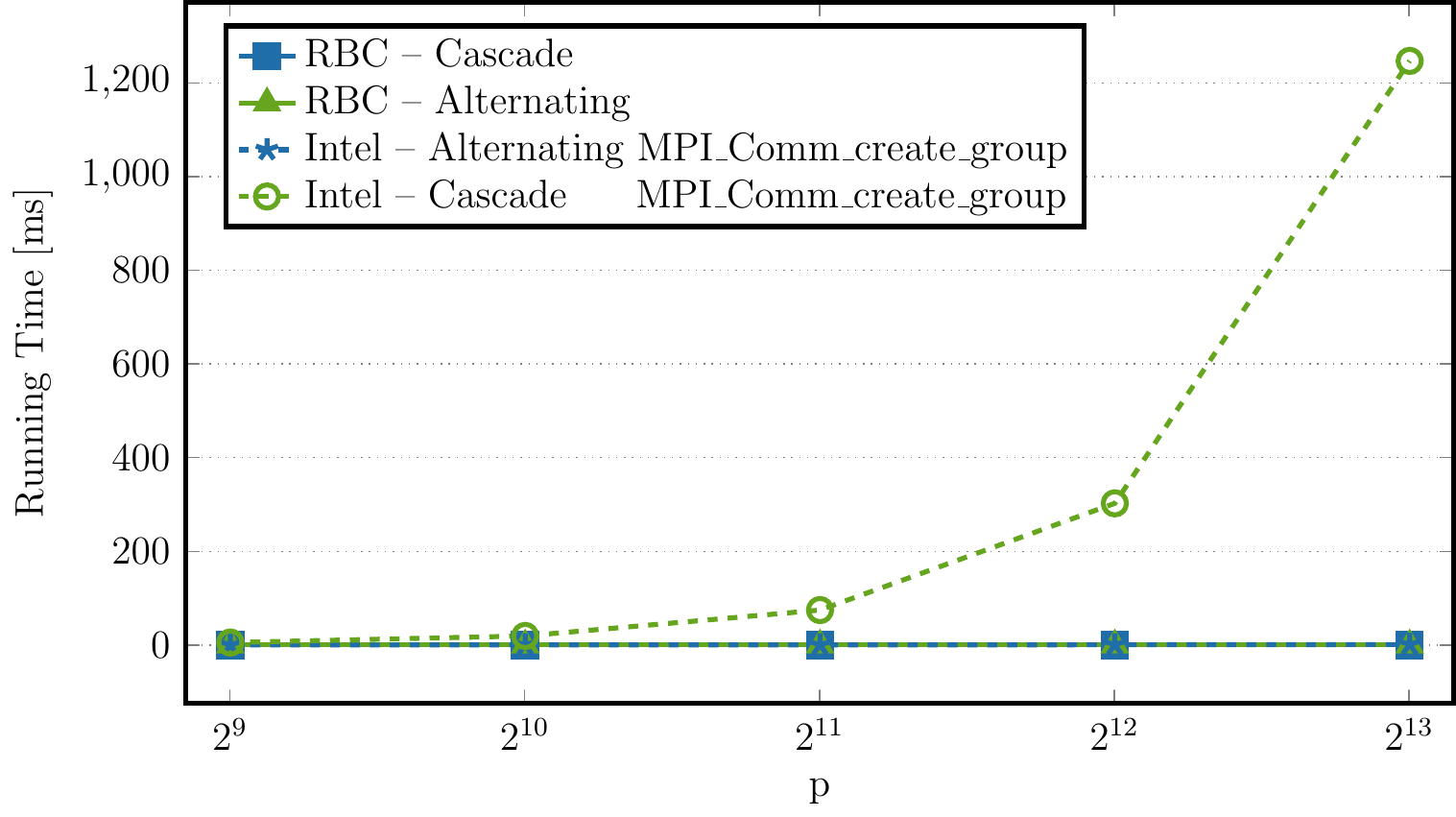}
  \caption{
    Running times of splitting a communicator of $p$ processes into overlapping communicators of size $4$ with a cascaded schedule and an alternating schedule.
  }
  \label{fig:cascade}
\end{figure}

\emph{Overlapping communicators}.
When multiple processes create multiple communicators at the same time, they must decide which communicator will be created first.
A wrong schedule results in cascaded communicator creation or even deadlocks.
In this case processes will delay communicator construction until other communicators have been created.
In our experiment we split a communicator of $p$ processes into communicators containing processes $\myset{0}{3}$,  $\myset{3}{6}$, $\myset{6}{9}$, and so on.
Note that processes $3, 6, 9$ and so on create two communicators -- one communicator contains processes to the left, one communicator contains processes to the right.
In the \emph{cascaded schedule}, processes which will be part of two communicators always create the left communicator first and the right communicator second.
In the \emph{alternating schedule}, every other process which will be part of two communicators creates the left communicator first, the other processes which will be part of two communicators create the right communicator first.
The alternating schedule avoids cascades but can be expensive as processes which create two communicators need global view.
Figure~\ref{fig:cascade} gives running times of cascaded and alternating splitting with Intel MPI and the RBC library.
The running time for communicator creation with our library is negligible.
There is almost no difference in the running time between cascaded and alternating scheduling as both operations are executed just locally.
However, the running time of cascaded communicator creation with Intel MPI becomes extremely slow for a large number of processes as the communicator construction of one group prohibits the communicator construction of other groups.
We do not give running times with IBM MPI as the function \texttt{MPI\_Comm\_create\_group} of IBM MPI is slower than its counterpart in Intel MPI by multiple orders of magnitude (see Figure~\ref{fig:split}).

\begin{figure}[t]
	\centering
	\includegraphics[width=\linewidth]{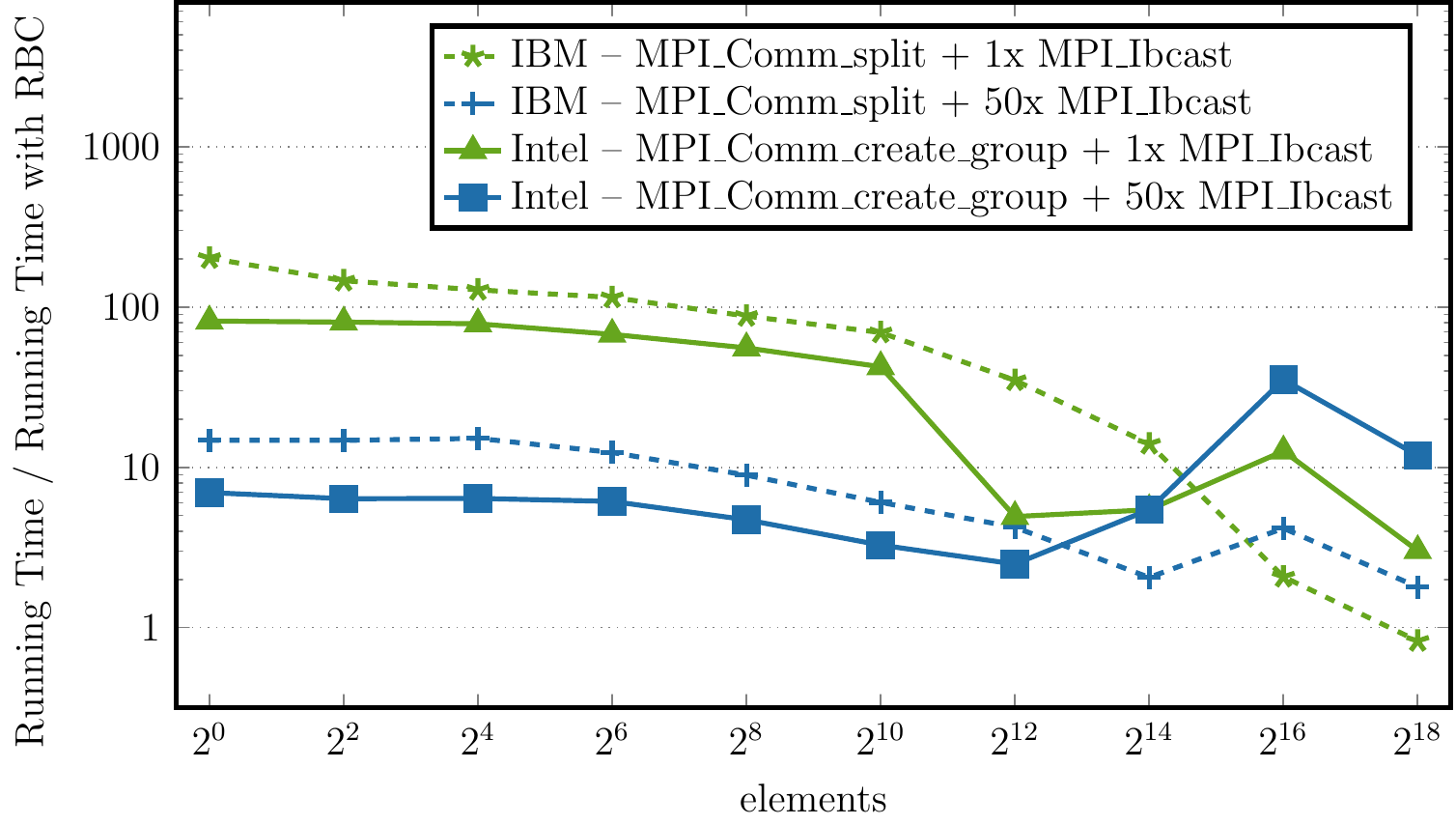}
	\caption{Running time ratios of MPI to RBC for broadcast operations on a sub-range of $2^{14}$ processes of an initial communicator of $2^{15}$ processes.}
        \label{fig:range bcast}
\end{figure}

\emph{Range-based collective}. The next benchmark performs the operation broadcast on a process range of a parent communicator.
We performed two experiments.
In the first experiment we split the parent communicator of size $p$ into a communicator of size $p/2$ and invoke the operation broadcast on the new communicator.
In the second experiment we also split the communicator once but invoke the operation broadcast $50$~times.
To split the parent communicator into an MPI communicator, we used the MPI function which performed best in the previous microbenchmark.
For Intel MPI (IBM MPI) we used the operation \texttt{MPI\_Comm\_create\_group} (\texttt{MPI\_Comm\_split}) to split the parent communicator.
If the RBC library is used, the communicators are split by invoking \texttt{rbc::Split\_RBC\_Comm}.
Figure~\ref{fig:range bcast} gives running time ratios of MPI to RBC for both experiments on $2^{15}$ processes.
For a moderate number $n$ of elements on the root process, $n\leq 2^{10}$, our library performs a single range-based broadcast faster than the Intel MPI (IBM MPI) library by a factor of $42.5$ to $81.8$ ($69.4$ to $202$).
For the same input sizes, our library performs $50$ range-based broadcasts faster than the Intel MPI (IBM MPI) library by a factor of $3.2$ to $7$ ($6$ to $14.7$).
For large inputs, the running times of IBM MPI converge to the running times of the RBC library.
The running times of Intel MPI fluctuate for large $n$.
In conclusion RBC performs nonblocking broadcast operations on a sub-range of processes whereas MPI must split the communicator with a blocking operation before a (non)blocking broadcast operation can be invoked.
In addition our library outperforms its competitors for almost all inputs as the time to create the MPI sub-communicators dominates for moderate input sizes.

\subsection{Sorting Benchmark}\label{sec:sorting benchmark}

\begin{figure}[t]
  \centering
  \includegraphics[width=\linewidth]{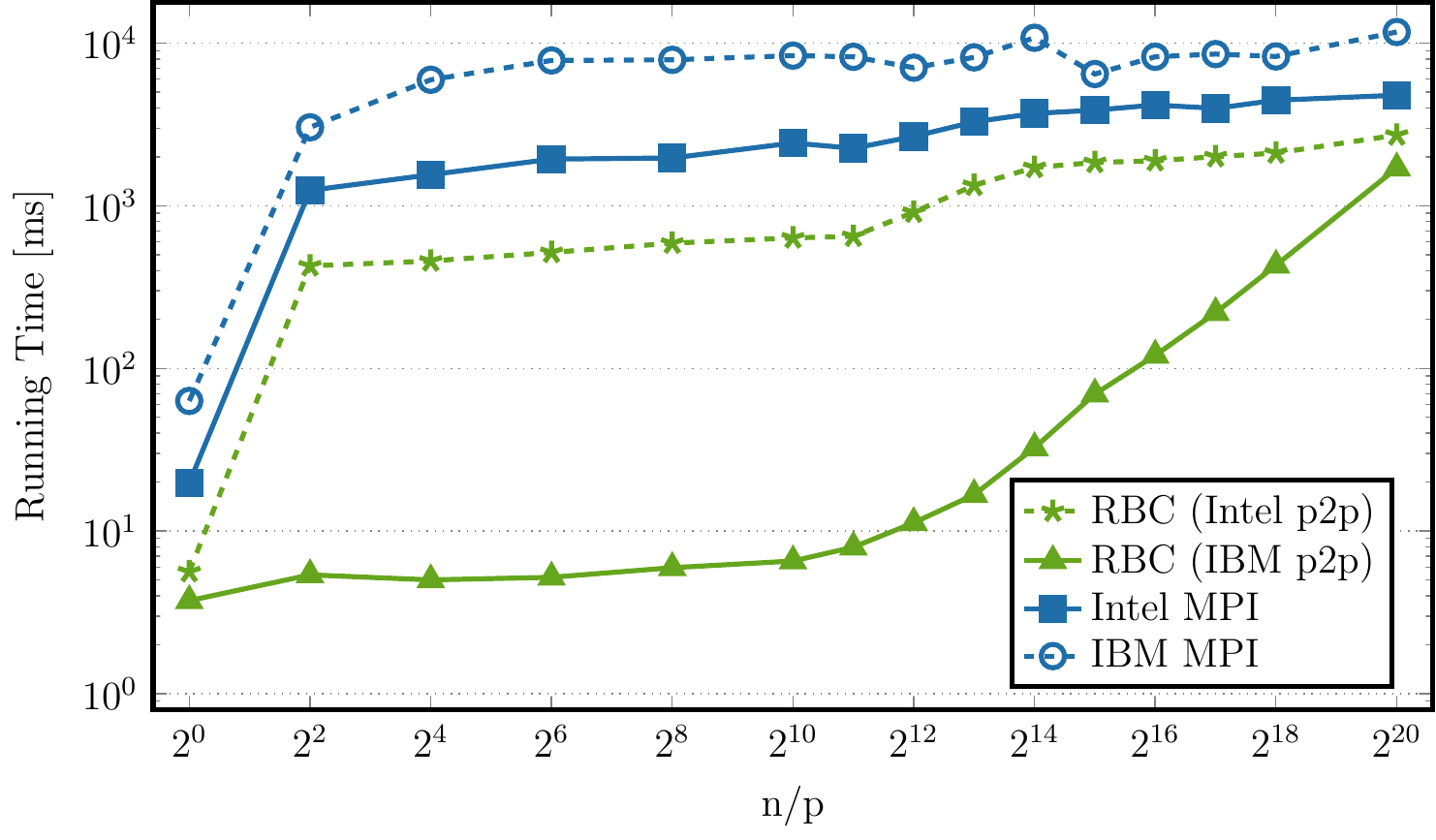}
  \caption{Running times of JQuick on $2^{15}$ cores with IBM MPI, Intel MPI, and RBC on top of IBM MPI and Intel MPI.}
  \label{fig:sorting with mpi ibm}
\end{figure}

In Figure~\ref{fig:sorting with mpi ibm} we present running times of JQuick implemented with RBC, native Intel MPI, and native IBM MPI.
We executed the experiments with our RBC library on top of IBM MPI as well as on top of Intel MPI.
All implementations exploit an alternating splitting schedule to avoid cascaded communicator construction.
In our alternating schedule every other janus process splits the left group first and the remaining janus processes split the right group first.
For $n/p=1$ no janus processes occur but JQuick with RBC already outperforms JQuick with native MPI by a factor of $3.5$ (Intel) to $16.9$ (IBM).
For moderate inputs, i.e., $1 < n/p \leq 2^{10}$, JQuick with RBC on top of IBM MPI outperforms JQuick with native IBM MPI by a factor of more than $1282$, even though no cascades occur.
For larger inputs, the running time of both implementations converge as the time of communicator construction becomes dominated by the actual algorithm.
JQuick with RBC on top of Intel MPI and JQuick with native Intel MPI are significantly slower for any input size.
This is caused by immense fluctuations.
We were able to avoid these fluctuations for $p\leq 4096$ by disabling dynamic connections (\texttt{I\_MPI\_DYNAMIC\_CONNECTION=no}).
However, this option did not work for large $p$.
Still, we see that the version with RBC is up to a factor of $3.8$ faster than the version with native Intel MPI for moderate inputs.
Experiments with a cascaded schedule showed that the running time of JQuick with RBC remains the same.
However, the versions with native MPI become even slower by multiple orders of magnitude for moderate inputs.

\section{Conclusion and Future Work}

We have shown how practical algorithms benefit from lightweight communicators.
We proposed RBC, a library which creates range-based communicators without communication in constant time.
The construction of our communicators come with almost no cost while providing nonblocking collective operations as well as point-to-point communication on a sub-range of processes.
This offers completely new ways to implement flexible algorithms efficiently.
We validated the performance of our communication routines in microbenchmarks as well as their applicability to sorting algorithms.
It would be interesting to apply RBC to other divide-and-conquer algorithms such as QuickHull~\cite{eddy1977new} and delaunay triangulation~\cite{funke2017parallel}.
RBC is even interesting from a theoretical side as recursive algorithms with polylogarithmic running time can now be implemented without communicator construction which takes time linear to the number of processes.

We propose and discuss a practical interface and implementation to integrate nonblocking (range-based) communicator creation into MPI.
We see this interface as an interesting candidate for the MPI standard.
Future work should include implementations of this interface into open-source MPI libraries to show their feasibility.

\iffullpaper
\section*{Acknowledgment}

The authors gratefully acknowledge the Gauss Centre for Supercomputing e.V. (\href{www.gauss-centre.eu}{www.gauss-centre.eu}) for funding this project by providing computing time on the GCS Supercomputer SuperMUC at Leibniz Supercomputing Centre (LRZ, \href{www.lrz.de}{www.lrz.de}).
This research was partially supported by the DFG project SA 933/11-1.
Additionally, we would like to thank Tobias Heuer
for a first prototypical implementation of JQuick.
\fi

\balance

\bibliographystyle{IEEEtran}
\bibliography{bibtex.bib}

\iffullpaper
\newpage

\onecolumn
\section{Appendix}

\begin{figure}[h]
	\centering
	\vspace*{20pt}
	\begin{subfigure}[t]{0.485\textwidth}
		\includegraphics[width=\linewidth]{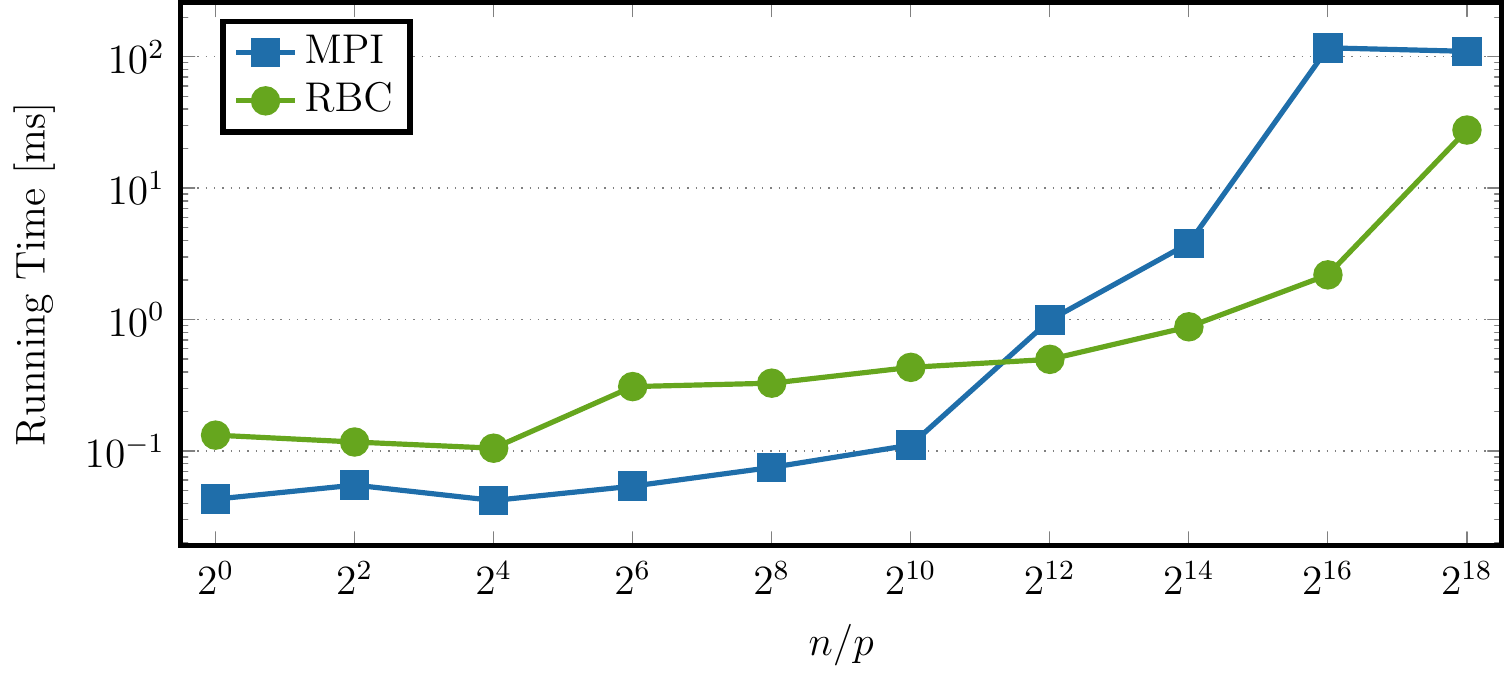}
		\caption{Broadcast with IBM MPI.}
	\end{subfigure}	
	\quad
	\begin{subfigure}[t]{0.485\textwidth}
		\includegraphics[width=\linewidth]{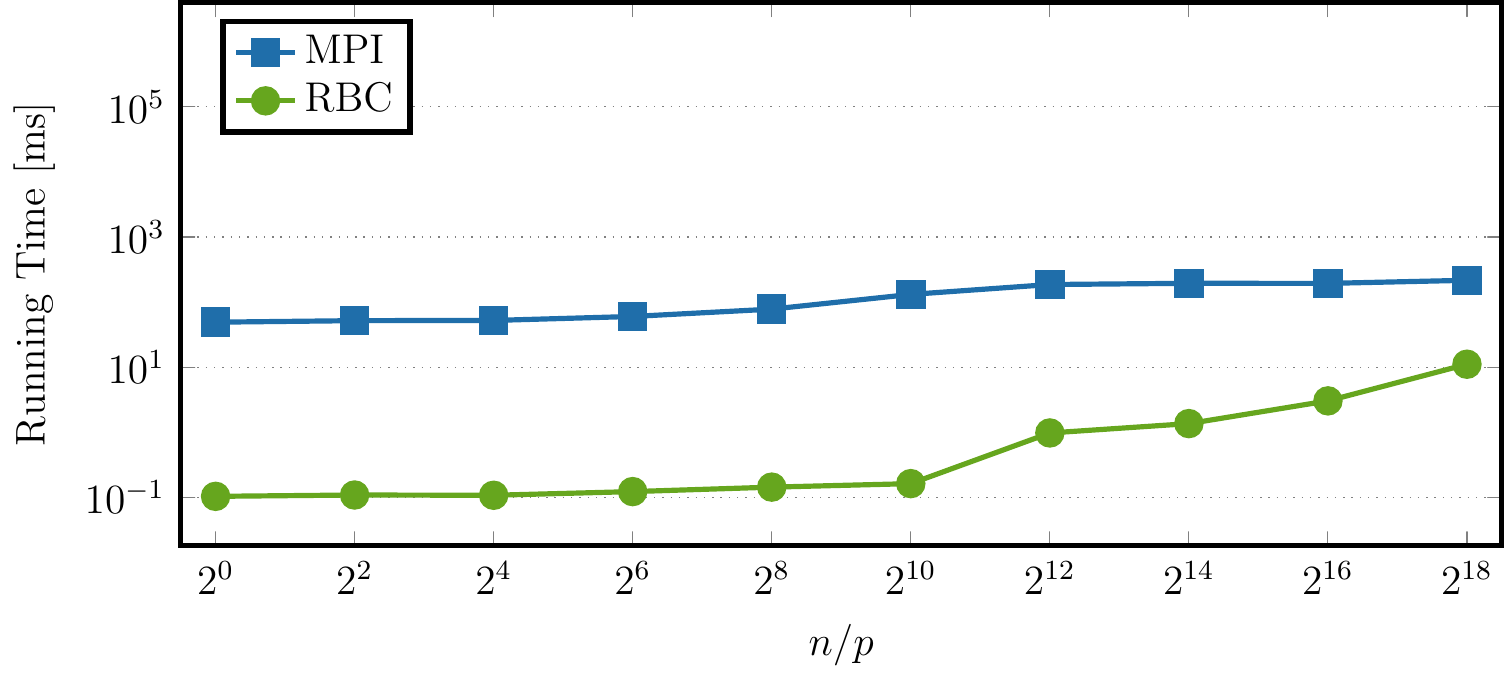}
		\caption{Broadcast with Intel MPI.}
	\end{subfigure}
	\vspace{10pt}
		
	\begin{subfigure}[t]{0.485\textwidth}
		\includegraphics[width=\linewidth]{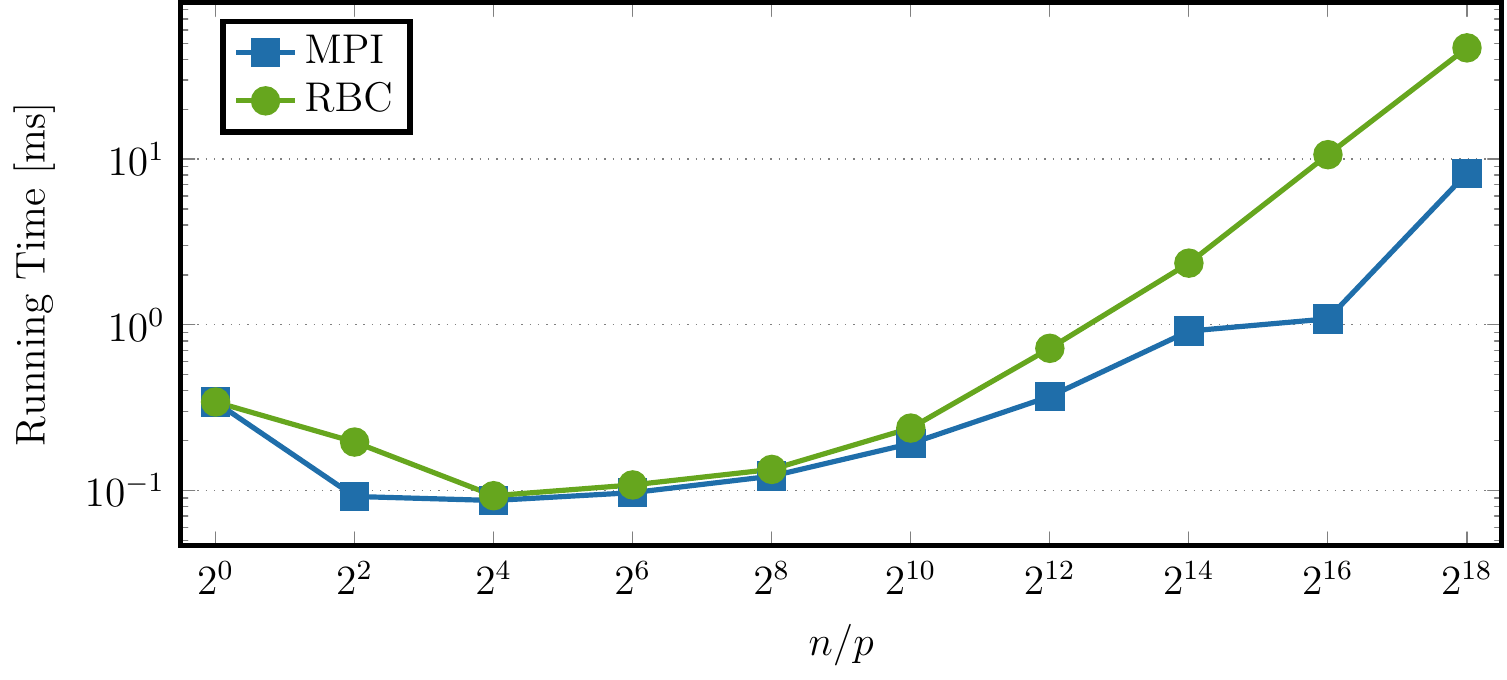}
		\caption{Reduce with IBM MPI.}
	\end{subfigure}	
	\quad
	\begin{subfigure}[t]{0.485\textwidth}
		\includegraphics[width=\linewidth]{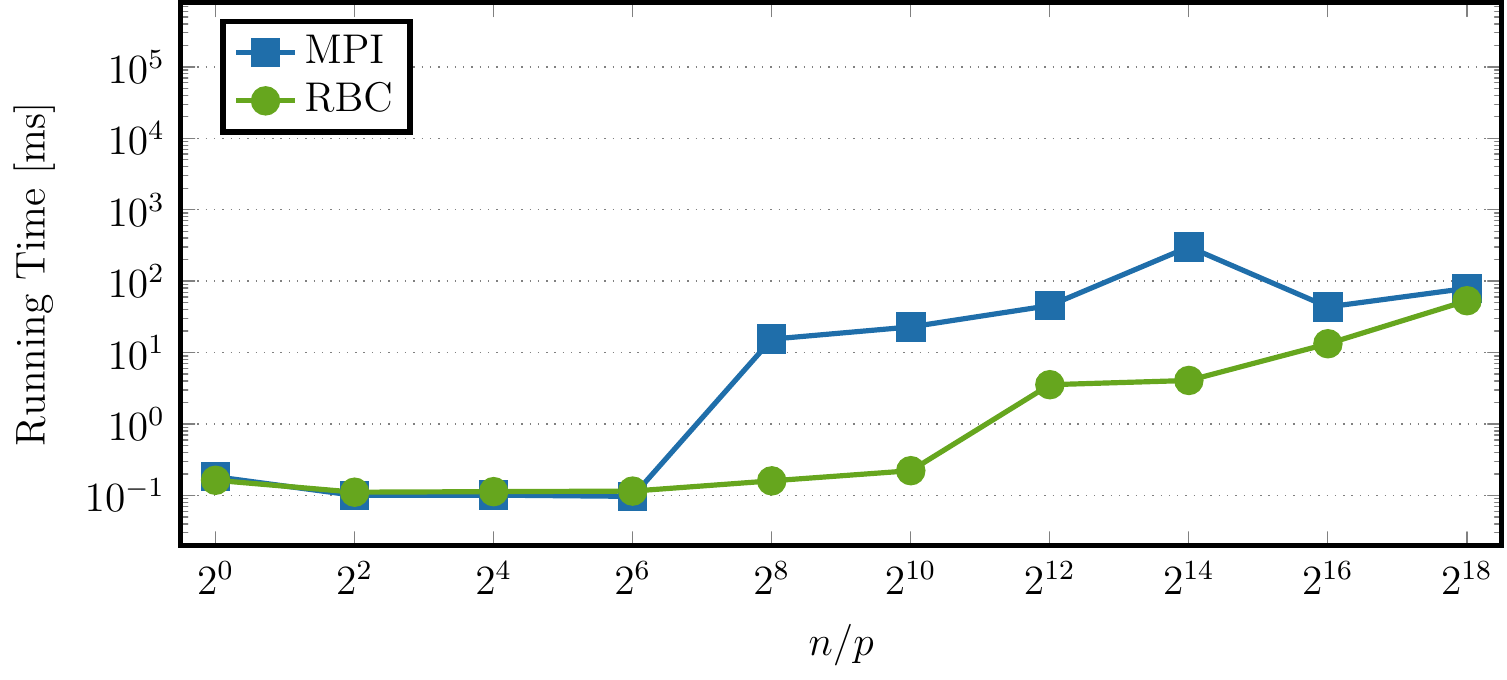}
		\caption{Reduce with Intel MPI.}
	\end{subfigure}
	\vspace{10pt}
	
	\begin{subfigure}[t]{0.485\textwidth}
		\includegraphics[width=\linewidth]{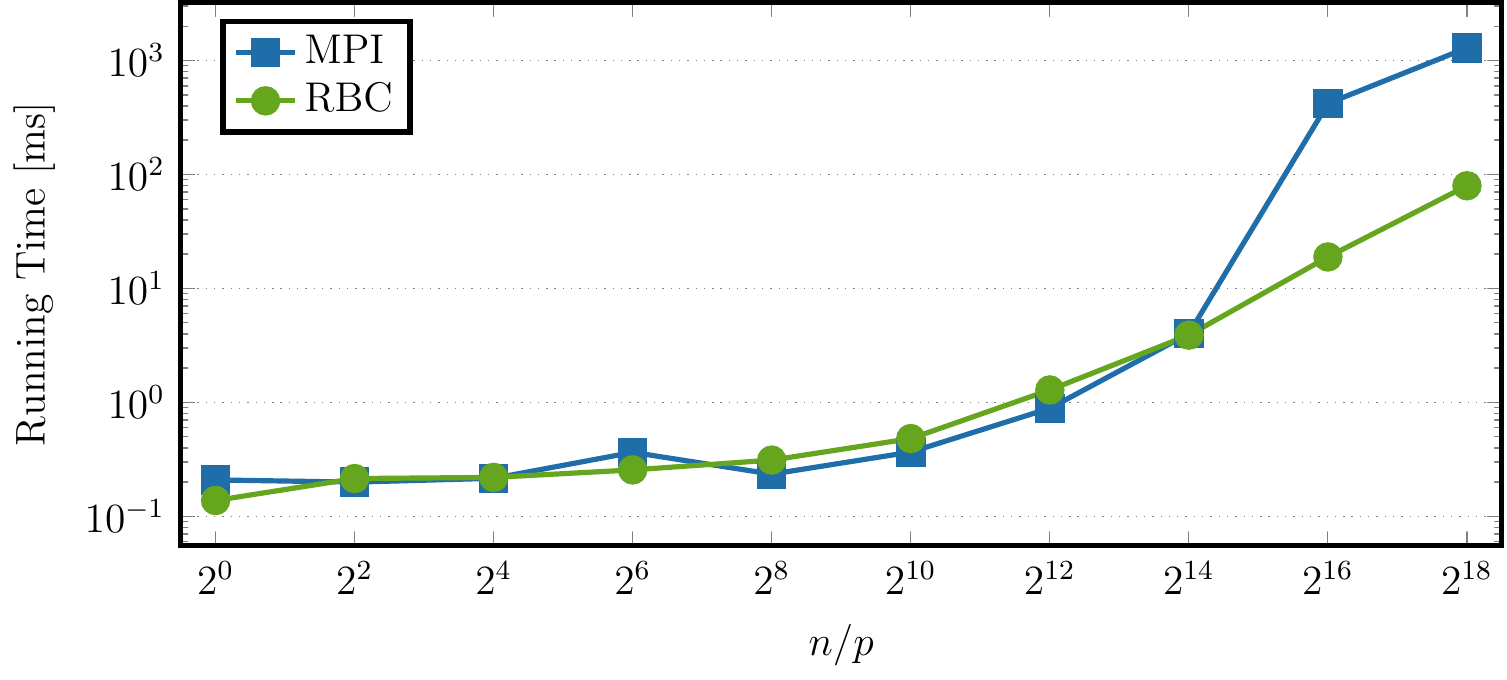}
		\caption{Scan with IBM MPI.}
	\end{subfigure}	
	\quad
	\begin{subfigure}[t]{0.485\textwidth}
		\includegraphics[width=\linewidth]{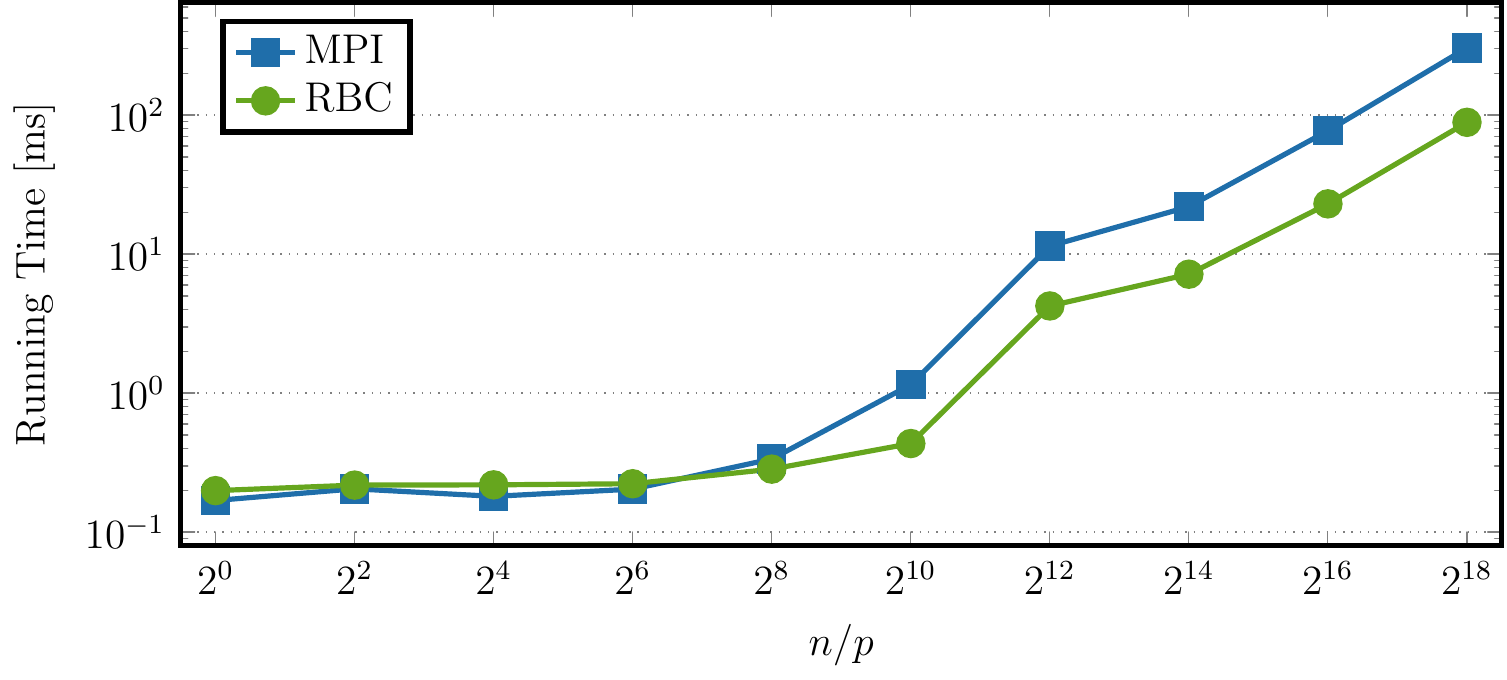}
		\caption{Scan with Intel MPI.}
	\end{subfigure}
	\vspace{10pt}
	
	\begin{subfigure}[t]{0.485\textwidth}
		\includegraphics[width=\linewidth]{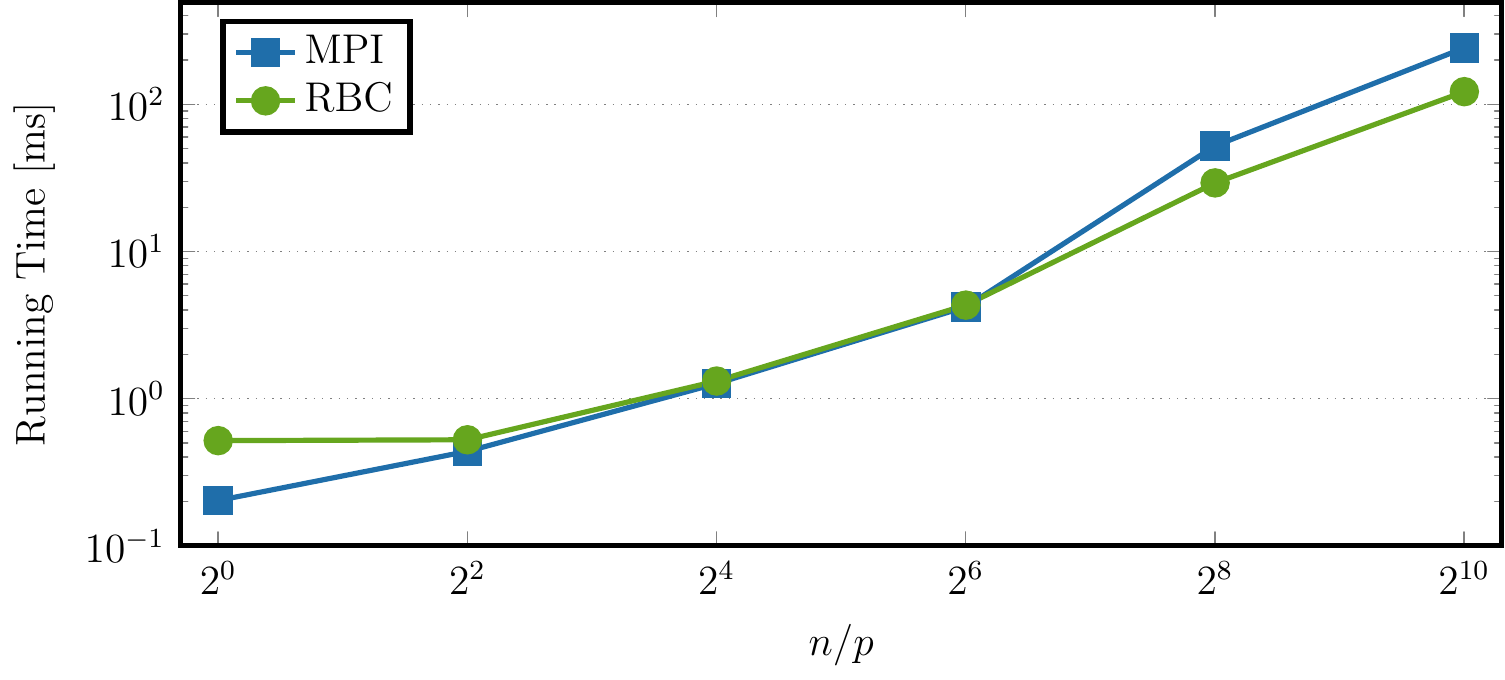}
		\caption{Gather with IBM MPI.}
	\end{subfigure}
	\quad
	\begin{subfigure}[t]{0.485\textwidth}
		\includegraphics[width=\linewidth]{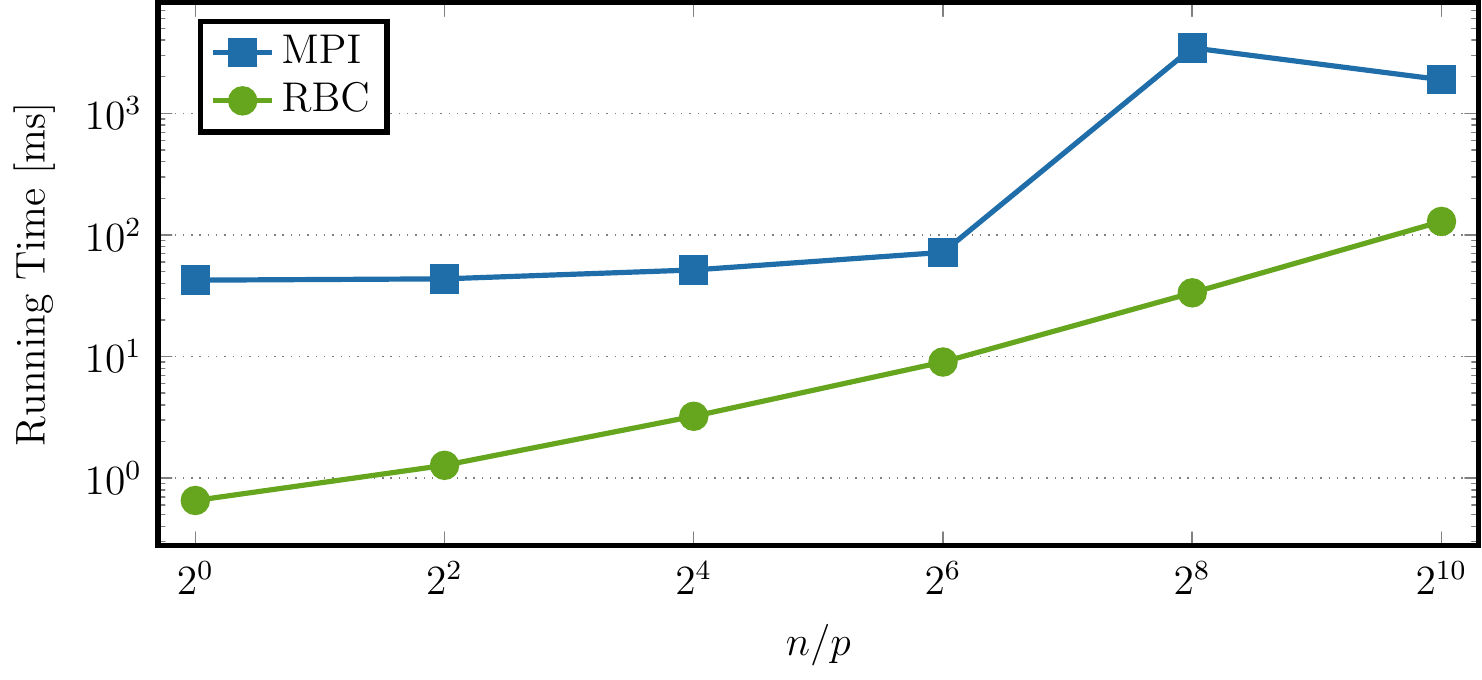}
		\caption{Gather with Intel MPI.}
	\end{subfigure}
	\caption{Running times of nonblocking collective operations on $2^{15}$ cores.}
        \label{fig:all colls}
\end{figure}
\fi

\end{document}